%% file: main.tex
\documentclass[a4paper,12pt]{article}

\usepackage{amsmath}
\usepackage{latexsym}
\usepackage{cite}
\usepackage{graphicx}
\usepackage[small,bf,hang]{caption}

\input{commands}

\begin{document}
\bibliographystyle{JHEP}

%TITLEPAGE
\renewcommand{\thefootnote}{\fnsymbol{footnote}}
\thispagestyle{empty}
\begin{titlepage}

%BUTP Nr.
\vspace*{-1.5cm}
\hfill \parbox{3.5cm}{hep-th/0301155\\TUW-03-03\\BUTP-2003/01}
\vspace*{0.2cm}

% Title
\begin{center}
  {\Large {\bf \hspace*{-0.2cm} SUSY Glue-Balls,\protect\vspace*{0.2cm} \\ 
      Dynamical Symmetry Breaking\protect\vspace*{0.3cm}\\
      and Non-Holomorphic Potentials.}}
      \vspace*{1cm} \\

% Authors
{\bf L. Bergamin\footnote{email: bergamin@tph.tuwien.ac.at, phone: +43  1 58801 13622, fax: +43 1 58801 13699}}\\
      Institute for Theoretical Physics \\
    Technical University of Vienna \\
    Wiedner Hauptstr.\ 8-10 \\
    A-1040 Vienna, Austria\\[3ex]
{\bf P. Minkowski\footnote{email: mink@itp.unibe.ch, phone: +41 31 631 8624,
    fax: +41 31 631 3821}}\\
    Institute for Theoretical Physics \\
    University of Bern \\
    Sidlerstrasse 5\\
    CH - 3012 Bern, Switzerland
   \vspace*{0.8cm} \\  

% Date
%\today

\vspace*{1cm}

% Abstract
\begin{abstract}
\noindent
We discuss the instability of the Veneziano-Yankielowicz effective action (or
its supersymmetric ground-state)\ with respect to  higher order
derivative terms. As such terms must be present in an effective action, the
V-Y action alone cannot describe the dynamics of SYM consistently. We
introduce an extension of this action, where all instabilities are removed by
means of a much richer structure of the K\"ahler potential. We demonstrate that the dominant contributions to the
effective potential are determined by the non-holomorphic part of the action
and we prove that the non-perturbative ground-state can be equipped with
stable dynamics. Making an expansion near the
resulting ground-state to second order in the derivatives  never leads back to the result by Veneziano and
Yankielowicz. As a consequence new dynamical effects arise, which are interpreted as the formation of
massive states in the boson sector (glueballs) and are accompanied by dynamical supersymmetry breaking. As this
regime of the dynamics is not captured by standard semi-classical analysis
(instantons etc.), our results do not contradict these calculations but
investigate the physics of the system beyond these approximations.

% \vspace{3mm}
% \noindent
% {\footnotesize {\it PACS:} 11.30.Qc; 11.30.Pb; 11.15.Kc \newline
% {\it keywords:} dynamical supersymmetry breaking; low
% energy approximations; SYM theory}
\end{abstract}
\end{center}

\end{titlepage}

\renewcommand{\thefootnote}{\arabic{footnote}}
\setcounter{footnote}{0}
\numberwithin{equation}{section}
%TEXT
%
% \maketitle

\input{introduction}
\input{vymodel}
\input{vytononlocal}
\input{construction}
\input{discussion}
\input{conclusions}

\section*{Acknowledgement}
One of us (L.B.)\ would like to thank E.\ Scheidegger for numerous interesting
discussions. This work has been supported by the Swiss National Science Foundation (SNF) and
the Austrian Science Foundation (FWF) project P-16030-N08.
%%
%%APPENDIX
%%
\appendix
\bibliography{biblio}
\end{document}

%% file: commands.tex
\newcommand{\cg}{C(G)}

\newcommand{\fdual}{\tilde{F}}

\newcommand{\cc}{^*}

\newcommand{\lomega}{\bra{\Omega}}

\newcommand{\lrpartial}{{\stackrel{\leftrightarrow}{\partial}\!}\mbox{\hspace{0.17ex}}}

\newcommand{\medsp}{\\[0.7ex]}

\newcommand{\romega}{\ket{\Omega}}

\newcommand{\veff}{V_{\mbox{\tiny eff}}}

\newcommand{\Lindex}[1]{\ensuremath{\smallindex{\mathcal{L}}{#1}}}
\newcommand{\Ltext}[1]{\ensuremath{\itindex{\mathcal{L}}{#1}}}

\newcommand{\bra}[1]{\langle #1 |}
\newcommand{\dega}{\ensuremath{^\dag}}

\newcommand{\diff}[1]{\mbox{d}#1}

\newcommand{\half}[1]{\ensuremath{\frac{#1}{2}}}
\newcommand{\intd}[1]{\int \!\! #1 \;}
\newcommand{\inv}[1]{\ensuremath{\frac{1}{#1}}}
\newcommand{\ket}[1]{| #1 \rangle}
\newcommand{\metr}[1][]{g_{\varphi \bar{\varphi} #1}}
\newcommand{\metrhat}[1][]{g_{\hat{\varphi} \Hat{\Bar{\varphi}} #1}}
\newcommand{\metrtilde}[1][]{\Tilde{g}_{\varphi \bar{\varphi} #1}}

\newcommand{\itindex}[2]{\ensuremath{#1_{\mbox{\scriptsize{\itshape #2}}}}} 
\newcommand{\smallindex}[2]{\ensuremath{#1_{\scriptscriptstyle{#2}}}}

\newcommand{\varfrac}[2][]{\frac{\delta #1}{\delta #2}}

\DeclareMathOperator{\tr}{Tr}
\DeclareMathOperator{\hc}{h.c.}

%% file: introduction.tex
\section{Introduction}
The low-energy dynamics of $N=1$ SYM theory are described in terms of a chiral
superfield $\Phi = \varphi + \theta \psi + \theta^2 F$. Its lowest
component $\varphi$ is the gaugino condensate while the highest one is the
scalar glue-ball operator. The effective action is given
by the most general Lagrangian of this superfield obeying all symmetries. It has
been shown in ref.\ \cite{bergamin02:1} that the most general Lagrangian of a
chiral superfield $\Psi$ does not consist of polynomials of this field alone,
but of polynomials in infinitely many chiral fields $\Psi_n$, where
\begin{align}
  \label{eq:psidefintro}
  \Psi_0 &= \Psi\ , & \Psi_n &= \bar{D}^2 \bar{\Psi}_{n-1}\ .
\end{align}
The complete (non-local) action can then be written as
\begin{equation}
  \label{eq:nonlocalintro}
    \mathcal{L} = \intd{\diff{^4x}} \biggl( \intd{\diff{^4 \theta}} K(\Psi_n,
  \bar{\Psi}_n) - \bigl(\intd{\diff{^2 \theta}} W(\Psi_n) + \hc \bigr) \biggr)\ ,
\end{equation}
where both $K$ and $W$ are polynomial functions, i.e.\ before using the
constraint \eqref{eq:psidefintro} they describe the standard K\"{a}hler- and
superpotential, respectively. All higher derivatives arise from the constraint
\eqref{eq:psidefintro}.
In certain cases
stability requirements force us to add a term
of the form $\bar{\Psi} \Psi \partial_\mu \Psi \partial^\mu \bar{\Psi}$
transgressing the form \eqref{eq:nonlocalintro}. This leads to an alternative
description of the model, namely
\begin{equation}
  \label{eq:intro1}
  \mathcal{L} = \intd{\diff{^4x}} \biggl( \intd{\diff{^4 \theta}} A(\Psi_0,\Psi_1 , \bar{\Psi}_0, \bar{\Psi}_1) + \bigl(
  \intd{\diff{^2 \theta}} H(\Psi_0) + \hc \bigr) \biggr) \ ,
\end{equation}
where $A(\Psi_0,\Psi_1 , \bar{\Psi}_0, \bar{\Psi}_1)$ and $H(\Psi_0)$ are no
longer polynomial functions in the superfields but also contain space-time
derivatives. This Lagrangian \eqref{eq:intro1} describes the most general form
of the effective action.

In this paper we construct a specific example of
\eqref{eq:nonlocalintro}/\eqref{eq:intro1} obeying all symmetries and
stability requirements of an effective action for $N=1$ SYM\footnote{Some
  steps of this construction had been performed in ref.\ \cite{Shore:1983kh}
  already, but a detailed analysis of the system had not been given therein.}. In this action
the two fields $\Psi_0$ and $\Psi_1$ are given in terms of the low-energy
field $\Phi$ as
\begin{align}
\label{eq:intro2}
  \Psi_0 &= (\Phi)^{\inv{3}}\ , & \Psi_1 &= \bar{D}^2 \bar{\Psi}_0\ .
\end{align}
The reason for the fractional power in equation \eqref{eq:intro2} is discussed
in section \ref{sec:constructionone}.

We show that the inclusion of $\Psi_1$ leads to important modifications
compared to the restricted action of \cite{veneziano82}, which is polynomial
in $\Psi_0$ alone:
\begin{equation}
  \label{eq:VYintro}
  \Ltext{V-Y} = \intd{\diff{^4x}} \biggl( \intd{\diff{^4 \theta}} K(\Psi_0,
  \bar{\Psi}_0) - \bigl(\intd{\diff{^2 \theta}} W(\Psi_0) + \hc \bigr) \biggr)
\end{equation}
These modifications are a consequence new terms appearing in the
non-holomorphic part of the action: The inclusion of $\Psi_1$ allows terms in
any power of $F$ while \eqref{eq:VYintro} is restricted to terms linear and
quadratic in this field. Therefore the (physically relevant) body of the
K\"{a}hler potential (in particular the K\"{a}hler metric $\metr$) is a
function of $\varphi$ \emph{and} $F$ in
\eqref{eq:nonlocalintro}/\eqref{eq:intro1}: $K = K(\varphi, F; \bar{\varphi},
\bar{F})$. This essential dependence on $F$ allows to define a (effective) potential
with massive particles and all relevant couplings even for $W \equiv 0$. Thus
the spectrum of the theory is not defined by the superpotential alone.
The generalized form
\eqref{eq:nonlocalintro}/\eqref{eq:intro1} changes the physical spectrum
completely compared to \eqref{eq:VYintro}. Moreover our action can have any power of
derivatives on all fields (including $F$), while \eqref{eq:VYintro} has
no derivatives on $F$.

We will prove below that these two points must lead to an essentially
different behavior of the theory than \eqref{eq:VYintro} has. The action \eqref{eq:VYintro} can never be seen as an
approximation. Physical and mathematical requirements on the
behavior of an effective action tell us that solely the general models
\eqref{eq:nonlocalintro} or \eqref{eq:intro1} are acceptable as an ansatz of the
latter.  

The existence of higher order derivative terms and of an effective potential
independent of the superpotential has important consequences for the
holomorphic terms as well. Indeed, the
restriction of $H$ being a holomorphic function in the superfields
$\Psi_n$ is one of the most powerful restrictions from supersymmetry on the
general form of the action. If $\mathcal{L}$ is the fundamental Lagrangian,
holomorphicity leads to the non-renormalization theorems in perturbation
theory. But even in the non-perturbative region holomorphic dependence leads
to severe constraints on the behavior of the theory, summarized e.g.\ in
\cite{intriligator96,peskin97}. Besides purely
field-theoretical statements interesting
relations to string- and M-theory are based on holomorphic structure as well: Recent results have shown how to obtain the
superpotential $W(\Phi)$ of the action \eqref{eq:VYintro} from such models
\cite{Dijkgraaf:2002fc,Dijkgraaf:2002vw,Dijkgraaf:2002dh,Dijkgraaf:2002xd}.

Nevertheless, it is impossible to discuss the role of the superpotential
without taking into account the non-holomorphic terms. This is almost trivial
if the action is seen as a fundamental Lagrangian, e.g.\ as a
Wess-Zumino model, where as an example the mass of the particles is determined by the quadratic
part of the superpotential:
\begin{equation}
\label{eq:introd2}
  V_W = m \intd{\diff{^2 \theta}} \Phi \Phi + \hc = m (F \varphi - \psi
  \psi) + \bar{m} ( \bar{F} \bar{\varphi} - \bar{\psi} \bar{\psi})
\end{equation}
Obviously this expression gives a mass $m$ to the spinor $\psi$, but as it
stands it does not define a mass for $\varphi$. This is obtained by writing
down the complete potential including terms from the non-holomorphic function
$K(\Phi, \bar{\Phi}) = \bar{\Phi} \Phi$ leading to
\begin{equation}
  \label{eq:introd3}
  V = - \bar{F} F +  m (F \varphi - \psi
  \psi) + \bar{m} ( \bar{F} \bar{\varphi} - \bar{\psi} \bar{\psi})\ .
\end{equation}
Completion of the square in the above equation yields the supersymmetric
spectrum. The potential in $F$ reduces to $V_F = - \bar{F} F$ and determining
$F$ by demanding the functional derivative to vanish represents a valid
constraint, reducing the degrees of freedom. This procedure is allowed as $F$ is an
auxiliary field in this case, i.e.\ there are no space-time derivatives acting
on $F$.

In an effective action, which must be based on \eqref{eq:nonlocalintro} or \eqref{eq:intro1},
the above procedure is impossible as $F$ is no longer an auxiliary
field. At this point the combination of the two modifications mentioned above
is important: Due to the derivatives acting on $F$ the potential of
\eqref{eq:VYintro} $V_F \propto -\bar{F} F$ becomes unstable.  Thus additional
non-holomorphic terms have to define a stable minimum. This is possible in
\eqref{eq:nonlocalintro}/\eqref{eq:intro1} as the K\"{a}hler potential is a
function of $\varphi$ and $F$. This
leads to dominant contributions to the potential from this part of the action, while the
role of the superpotential is changed completely (see section
\ref{sec:vytononlocal} below). These effective actions
have several typically non-perturbative features that cannot be found in
classical or perturbative supersymmetry \cite{bergamin02:1}.

The role of the non-holomorphic parts of supersymmetric Lagrangians eludes
a systematic study that is possible in case of the holomorphic terms. But
as the dominant contributions to the effective  potential of the actions \eqref{eq:nonlocalintro}/\eqref{eq:intro1}
 stem from the non-holomorphic part, we find that the ground-state of
supersymmetric Yang-Mills theories is characterized by essentially
non-perturbative effects, which at the present time are not available for
semi-classical and/or perturbative calculations. These effects are closely
related to the dynamics of the glue-ball \cite{bergamin01}, which --in
contrast to chiral symmetry breaking and quark-confinement-- is difficult to
understand in a semi-classical picture of non-supersymmetric QCD as well. In
addition dynamical supersymmetry breaking (DSB) occurs. This form of DSB
cannot be understood by semi-classical concepts as instanton-induced DSB and
defies perturbative and semi-classical arguments about the absence of
DSB in SYM \cite{bergamin01,bergamin01:2}.

We already stress at this point
that these conclusions are completely independent of the role of $F$ once the
true ground-state is determined. Indeed, having found the correct ground-state we may
arrive at the conclusion that the residual dynamics of $F$ are actually suppressed and
may thus be integrated out. This does \emph{not} lead back to the
interpretation of $F$ as an auxiliary field in the sense of
\eqref{eq:introd3}, but is a conclusion that applies in an expansion around
the correct ground-state only, where $F$ is \emph{never} an auxiliary
field. This seeming contradiction is the consequence of approximations versus
exact mathematical statements: Integrating out a field is a question of
correct approximation (or correct counting of orders of the scale $\Lambda$),
but to qualify a certain field as auxiliary is an exact mathematical
statement, where sub-leading effects in $\Lambda$ \emph{do matter} (cf.\ the
discussion of section \ref{sec:vytononlocal} and of reference \cite{bergamin02:1}).

The paper is organized as follows: The derivation of the Lagrangian by
Veneziano and Yankielowicz is shortly reviewed in section \ref{sec:vylag}, the
necessity of a generalization of this model  demonstrated in
section \ref{sec:vytononlocal}. In section \ref{sec:construction} we construct
the generalized model and discuss its fundamental properties and in section
\ref{sec:DSB} we relate these results to independent knowledge about
non-perturbative SYM theories. Finally we draw our conclusions in section \ref{sec:conclusion}.

%%% Local Variables: 
%%% mode: latex
%%% TeX-master: "main"
%%% End: 

%% file: vymodel.tex
\section{The Veneziano-Yankielowicz Lagrangian}
\label{sec:vylag}
The effective description of $N=1$ SYM theories goes back to the work by
Veneziano and Yankielowicz \cite{veneziano82}. The authors showed that it is
possible to obtain a low energy description of the theory by means of an
effective Lagrangian in terms of the anomaly multiplet (also referred to as
Lagrangian or glue-ball multiplet)
\begin{equation}
\label{eq:Phidef}
\Phi = \inv{8 \cg} \tr W^\alpha W_\alpha = \varphi + \theta \psi + \theta^2 F
\end{equation}
In the present paper we extend this Lagrangian by adding higher
derivative terms and we show that this procedure changes the role of the local
part of \cite{veneziano82} in a non-trivial way. To make our statements more
transparent we shortly review the arguments leading to the Lagrangian of
ref.\ \cite{veneziano82}.

It is most natural to assume that the low-energy dynamics of confined SYM
theories must be described in terms of the anomaly multiplet $\Phi$. Indeed,
its lowest component is the gaugino-condensate, the spinor represents the
Goldstino if supersymmetry breaks dynamically and its highest component
contains the operators $\tr F_{\mu \nu} F^{\mu \nu}$ and $\tr F_{\mu \nu}
\fdual^{\mu \nu}$ used to represent the anomaly of the $R$-current and
scale invariance. Also, this multiplet has a definite interpretation in the
underlying symmetry structure as it appears in the anomaly of the
supercurrent\footnote{This supercurrent in equation \eqref{eq:anomcurrent} is
  not to be confused with the local current of the supercharge $Q_\alpha$.}
\begin{equation}
  \label{eq:anomcurrent}
  \bar{D}^{\dot{\alpha}} V_{\alpha \dot{\alpha}} = \frac{2}{3 \cg}
  \frac{\beta(g)}{g^3} D_\alpha \Phi \ .
\end{equation}
Thus we can expect that the multiplet $\Phi$ is defined not only in perturbation
theory but
also in the non-perturbative region. The effective description is obtained by
writing down the most general Lagrangian in terms of $\Phi$ having the correct
symmetry properties. $\Phi$ is gauge-invariant by construction and
supersymmetry is realized linearly as $\Phi$ is a chiral superfield. Thus the
most general Lagrangian has the form
\begin{equation}
  \label{eq:localkahler}
    \mathcal{L} = \intd{\diff{^4x}} \biggl( \intd{\diff{^4 \theta}} K(\Phi,
  \bar{\Phi}) - \bigl(\intd{\diff{^2 \theta}} W(\Phi) + \hc \bigr) \biggr)\ .
\end{equation}
The remaining symmetries are the anomalous $R$-current, scale invariance and
special supersymmetry transformations ($S$-supersymmetry). Correct
anomaly cancellation is obtained by a term $\propto \Phi \log(z \Phi/\Lambda^3)$
with $\Lambda$ being the scale of the theory
\cite{veneziano82,burgess95}. Adding up all invariant terms that are
polynomial in the field $\Phi$, we
arrive at the Veneziano-Yankielowicz Lagrangian
\begin{equation}
  \label{eq:vylagrangian}
  \Ltext{V-Y} = \intd{\diff{^4x}}  \biggl( \intd{\diff{^4 \theta}} 9 a (\bar{\Phi} \Phi)^{\inv{3}} -
  \Bigl(\intd{\diff{^2 \theta}} \frac{2 \beta(g)}{9 \cg g^3} \bigl( \Phi \log
  \frac{z \Phi}{\Lambda^3} - \Phi \bigr)  + \hc \Bigr) \biggr) \ .
\end{equation}
Symmetries determine the Lagrangian up to three dimensionless constants $a$, $z$ and the renormalized coupling constant of the fundamental theory,
$g$. Performing the integrals over Grassman variables yields ($2 \beta /(9 \cg g^3)
= c$) 
\begin{multline}
\label{eq:vykahler}
  \intd{\diff{^4 \theta}} 9 (\bar{\Phi} \Phi)^{\inv{3}} =\medsp
  = (\bar{\varphi} \varphi)^{-\frac{2}{3}} \bigl(\partial_\mu \bar{\varphi}
  \partial^\mu \varphi + \half{i} \psi \sigma^\mu \lrpartial_\mu \bar{\psi} +
  \bar{F} F \bigr) - \frac{i}{9} \psi \sigma^\mu \bar{\psi} (\bar{\varphi}
  \varphi)^{-\frac{5}{3}} (\varphi \lrpartial_\mu \bar{\varphi}) \medsp
  + \inv{9} (\bar{\varphi} \varphi)^{-\frac{5}{3}} (\psi \psi) (\bar{\psi} \bar{\psi}) +
  \inv{3}  (\varphi^{-\frac{2}{3}} \bar{\varphi}^{-\frac{5}{3}} F \bar{\psi}
  \bar{\psi} + \hc )
\end{multline}
and
\begin{equation}
  \label{eq:vysuperpot}
  \intd{\diff{^2 \theta}} c \bigl(\Phi \log
  \frac{z \Phi}{\Lambda^3} - \Phi \bigr) = c \bigl( F \log \frac{z
  \varphi}{\Lambda^3} - \inv{2 \varphi} \psi \psi \bigr)\ .
\end{equation}
At this point Veneziano and Yankielowicz treat $F$ as an auxiliary field and
eliminate it by means of its
equation of motion (in accordance with \eqref{eq:introd3}). Indeed, varying \eqref{eq:vylagrangian} with respect to
$\bar{F}$ 
\begin{equation}
  \label{eq:vyelimination}
  F = - \inv{3} \frac{\psi \psi}{\varphi} + \frac{(\bar{\varphi}
  \varphi)^{\frac{2}{3}}}{a} c \log \frac{\bar{z} \bar{\varphi}}{\Lambda^3}
\end{equation}
and inserting this back into the Lagrangian an expression completely
independent of $F$ --and thus independent of $\tr F_{\mu\nu} F^{\mu \nu}$ and
$\tr F_{\mu\nu} \fdual^{\mu \nu}$-- is obtained:
\begin{equation}
  \label{eq:vyonshell}
  \begin{split}
    \Ltext{V-Y} &= \intd{\diff{^4x}} a (\bar{\varphi} \varphi)^{-\frac{2}{3}} \bigl(\partial_\mu \bar{\varphi}
  \partial^\mu \varphi + \half{i} \psi \sigma^\mu \lrpartial_\mu \bar{\psi}
  \bigr) - \frac{i a}{9} \psi \sigma^\mu \bar{\psi} (\bar{\varphi}
  \varphi)^{-\frac{5}{3}} (\varphi \lrpartial_\mu \bar{\varphi}) \medsp
  &\quad - \frac{(\bar{\varphi} \varphi)^{\frac{2}{3}}}{a} c^2 \log \frac{z
  \varphi}{\Lambda^3} \log\frac{\bar{z}
  \bar{\varphi}}{\Lambda^3} + c \Bigl((\psi \psi) (\inv{3 \varphi} \log \frac{\bar{z}
  \bar{\varphi}}{\Lambda^3} + \inv{2 \varphi}) + \hc \Bigr)
  \end{split}
\end{equation}
After having eliminated the auxiliary field the chirally broken minimum with
$\varphi_0 = \Lambda^3/z$ is found. Notice again that this minimum is not
obtained from the superpotential \eqref{eq:vysuperpot} alone. The
non-holomorphic terms in $F$ from \eqref{eq:vykahler} play a crucial role to
arrive at this conclusion.

The complex number $z$ determines the
phase of $\varphi_0$. In this work we consider the action as a quantum
effective action obtained from a source-extension (discussed in section
\ref{sec:construction} below). Vacuum alignment then dynamically determines
the phase of $\varphi_0$: $\varphi_0 < 0$ and thus  $z$ must be negative. This follows from
PCAC analysis \cite{gell-mann68} or QCD effective actions \cite{gasser82}, in
more general situations a complete study of thermodynamical limits
\cite{minkowski90,bergamin01,bergamin01:2} may be important. Thus we set $z$
in the following to $z = -1$. Making an expansion around this minimum
and absorbing the constant in front of the kinetic term by a field
redefinition
\begin{align}
  \varphi' &= \frac{\sqrt{a}}{\Lambda^2} \varphi\ , & \psi' &=
  \frac{\sqrt{a}}{\Lambda^2} \psi\ ,
\end{align}
one finds the supersymmetric spectrum with $m = (c/a) \Lambda$ as masses for
both, $\varphi$ and $\psi$.

%%% Local Variables: 
%%% mode: latex
%%% TeX-master: "main"
%%% End: 

%% file: vytononlocal.tex
\section{From Local to Non-Local Actions}
\label{sec:vytononlocal}
It is of utmost importance to observe that all conclusions from the
work by Veneziano and Yankielowicz --unbroken supersymmetry, confining vacuum
state with spectrum, positivity property of the potential-- can be drawn \emph{if and only if} the field $F$ is
auxiliary. We prove in this section that this statement must be an exact
mathematical requirement on the full theory (including all terms beyond the
approximation \eqref{eq:vylagrangian} or \eqref{eq:vyonshell}). Thus in any
enveloping theory, of which \eqref{eq:vyonshell} shall be an approximation,
$F$ remains an auxiliary field \emph{exactly}.

We will prove below that this restriction is untenable, as the
Veneziano-Yankielowicz Lagrangian must be extended by additional terms
promoting $F$ to an independent physical degree of freedom. Indeed, we cannot
expect that this action represents the complete dynamics at low energies: it
has derivative terms up to $\mathcal{O}(p^2)$ only, which is not acceptable
for an interacting theory. Generalizing \eqref{eq:vylagrangian} we write
\begin{equation}
  \label{eq:vygeneralization}
  \Ltext{SYM} (\Phi, \bar{\Phi}, \ldots) = \Ltext{V-Y}(\Phi, \bar{\Phi}) +
  \Delta \mathcal{L}(\Phi, \bar{\Phi}, \ldots)\ ,
\end{equation}
where the dots indicate that the full effective action could depend on additional
fields. The subsequent arguments of this section do not depend on the
approximation techniques as exemplified by the Wilsonian action, which is also
of the form \eqref{eq:vygeneralization}.

Whether $\Ltext{V-Y}$
is a good approximation of the whole theory or not depends on the details of
$\Delta \mathcal{L}$. Usually it is assumed that such an expansion up to second
order in the derivatives leads to a good approximation if all neglected terms
are suppressed by inverse powers of $\Lambda$. But stability/instability is not just a question
of the order of contributions from $\Delta \mathcal{L}$: If $\Ltext{V-Y}$
shall be an approximation of the whole theory, we thus have to
ensure that the fundamental behavior of the $F$ field (namely being auxiliary)
is not changed by $\Delta \mathcal{L}$.

The field $F$ has
mass dimension four and suppressing the dependence on the additional fields
present (or setting them to their ground-state values $\varphi = \varphi_0$
and $\psi \equiv 0$) the Lagrangian by  Veneziano and Yankielowicz remains of
the form \eqref{eq:introd3}
\begin{equation}
  \label{eq:vyFlag}
  \Lindex{F} = \frac{c_1}{\Lambda^4} \bar{F} F - d F - \bar{d} \bar{F}\ .
\end{equation}
Considering the stability of the system the linear terms are irrelevant. The potential in $F$ is not bounded
from below but instead has an absolute maximum, the typical behavior of an
auxiliary field. The supersymmetric spectrum obtained by Veneziano and
Yankielowicz is found by sitting on the top of the auxiliary-field potential,
which only after elimination of $F$ becomes the new minimum of the physical
potential.

We now add additional contributions from $\Delta \mathcal{L}$ to the
action. They should start at order $\mathcal{O}(\Lambda^{-6})$ and can include
derivative terms as well as modifications of the potential. Allowing additional derivative terms from $\Delta \mathcal{L}$, e.g.\
a new term
\begin{equation}
  \label{eq:vyFlag2}
  \Lindex{F} = \frac{|c_2|}{\Lambda^6} \partial_\mu F \partial^\mu \bar{F} + \frac{c_1}{\Lambda^4} \bar{F} F - d F - \bar{d} \bar{F}
\end{equation}
could appear, disqualifying the interpretation of $F$ as auxiliary field. Of course the kinetic
term in $F$ is formally suppressed compared to the kinetic
terms of $\varphi$ and $\psi$, which are of order
$\mathcal{O}(\Lambda^{-4})$. But at this point it is important to notice that
the stability of a system and the characterization of certain fields as auxiliary ones cannot be
restricted to some approximation but are \emph{exact mathematical
  statements}. If $c_2$ --or any higher order derivative in $F$-- is non-zero,
$F$ \emph{is no longer an auxiliary field}. Of course, this behavior has
nothing to do with supersymmetry, but is a simple statement about classical
mechanics: We put a ball on the top of a hill. When hitting it with an
arbitrarily small but finite momentum (i.e. some dynamics that can be suppressed by any
finite power of the typical scale (mass of the ball) of the system) this
inevitably leads to a dynamical revelation, i.e.\ the ball will roll down the
hill. A different conclusion is allowed if and only if the ball is an
``auxiliary ball''. This means that the dynamics have to vanish exactly \emph{in
the full theory} and we can't hit the ball at all, not even with an
infinitesimal momentum.

\begin{figure}[t]
\begin{center}
    \includegraphics[{scale=0.5}]{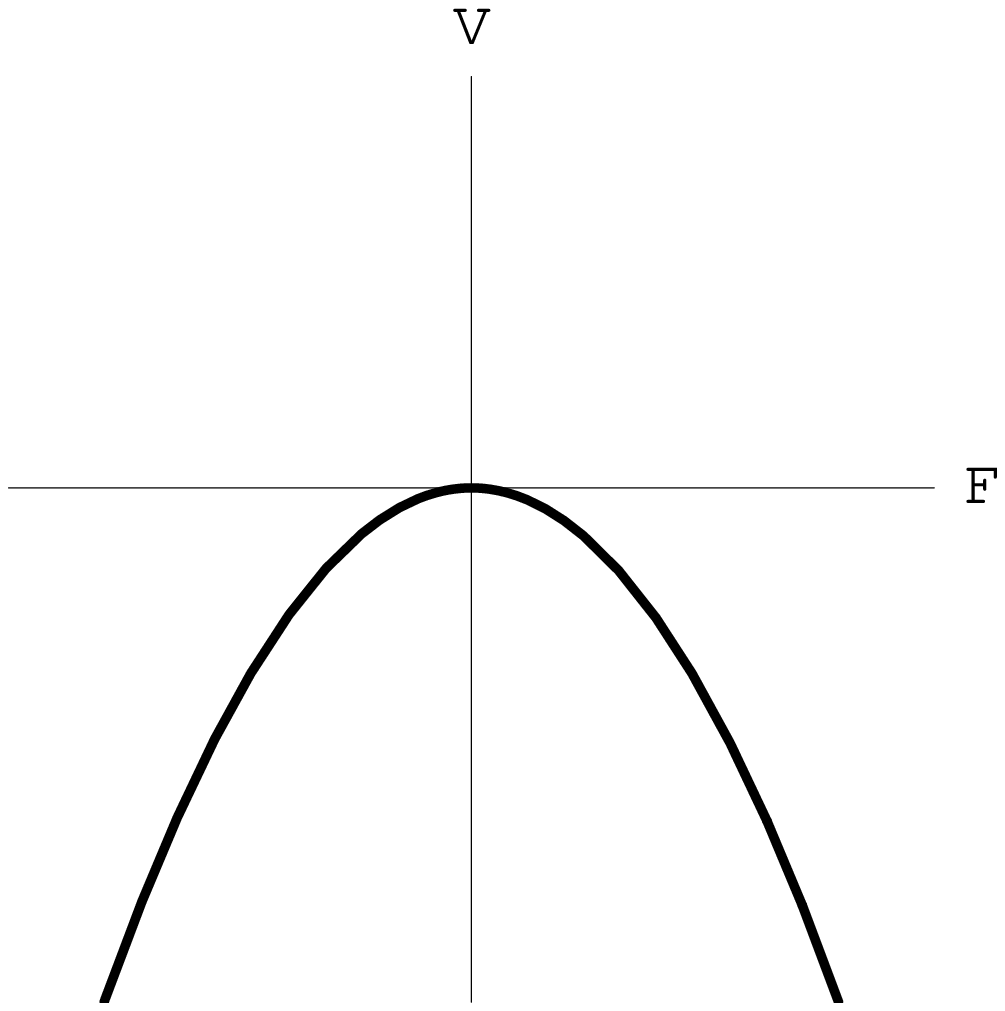} \hspace{3cm}
    \includegraphics[{scale=0.5}]{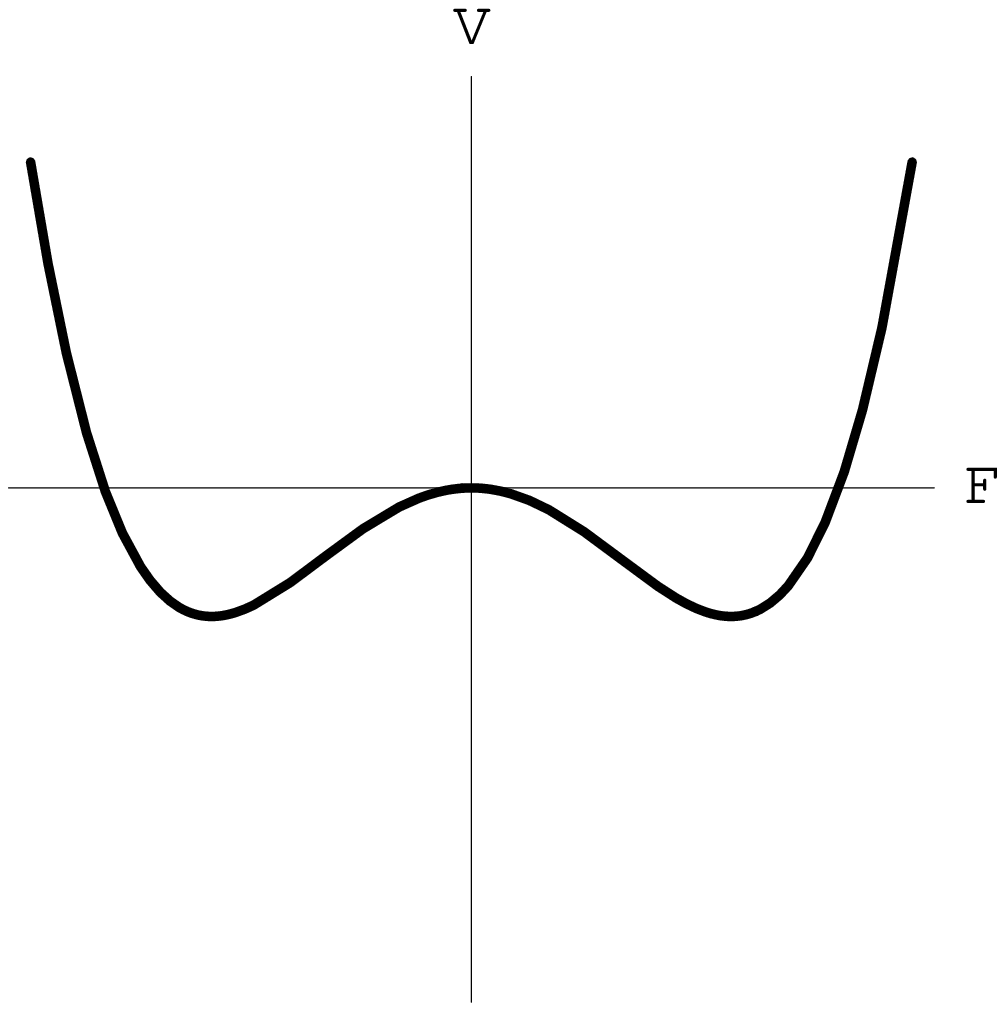}
\end{center}
    \caption{Additional terms to the potential of $F$ can turn the unstable
      potential from the Veneziano-Yankielowicz Lagrangian (left hand side) into
      an acceptable potential for dynamical $F$ (right hand side). If $F$ is
      auxiliary, the ground state is represented by the maximum of the
      Veneziano-Yankielowicz potential, if $F$ is dynamical the latter is
      represented by the minimum of the new potential.}
    \label{fig:potential}
\end{figure}

If $c_2$ (or any other term from $\Delta \mathcal{L}$ including derivatives on
$F$) is non-zero, the only way out is to provide for a finite bottom of the hill in \eqref{eq:vyFlag2} (cf.\ figure \ref{fig:potential}). If we assume that the action \eqref{eq:vyFlag2} is complete up to
order $\mathcal{O}(\Lambda^{-4})$ the simplest solution is
\begin{equation}
  \label{eq:vyFlag3}
  \Lindex{F} = \frac{|c_2|}{\Lambda^6} \partial_\mu F \partial^\mu \bar{F} +
  \frac{c_1}{\Lambda^4} \bar{F} F  - \frac{|c_3|}{\Lambda^{12}} (\bar{F} F)^2 - d F - \bar{d} \bar{F}\ .
\end{equation}
Now our ball will still roll down the hill, but at least this will stop when
it reaches the new minimum ($d = 0$ in the following)
\begin{equation}
  \label{eq:phi4min}
  F_0 = \Lambda^4 \sqrt{\frac{|c_1|}{2 |c_3|}}\ ,
\end{equation}
which is simply the minimum of a Mexican-hat potential. In this new minimum we
may still arrive at the conclusion that the dynamics of $F$ are suppressed and
that we may integrate this field out in a low-energy
approximation (notice that an application of this idea to supersymmetry won't
lead to a Goldstone mode from $F$). But this is now an approximation allowed
in the vicinity of the ground-state \eqref{eq:phi4min}, only. The correct ground-state is given by
\eqref{eq:phi4min} and the original interpretation of $F$ being an auxiliary
field (cf.\ \eqref{eq:vyFlag}) is \emph{never} an approximation to the true
behavior of the theory.

Equation \eqref{eq:vyFlag3} describes an effective Mexican-hat of some field
with mass dimension four. Abstracting from supersymmetry we once again
can make absolutely transparent what happens: The full dynamics of the system tell
us that the minimum of the potential is (up to a phase) at $F_0$ of equation
\eqref{eq:phi4min}. Sitting in this minimum we obtain the correct physical
spectrum of the theory and may integrate out some fields  if they are really
suppressed. Standing on the top of the hill however, power counting and
suppression arguments are absolutely fallacious. Else we would arrive at the
conclusion that both $c_2$ and $c_3$ are irrelevant effects and the theory
shrinks to the action \eqref{eq:vyFlag2}. Obviously this does not at all
describe the correct physics of a Mexican-hat like potential as the system is expanded around an unstable
point.

The question whether \eqref{eq:vyonshell} is a suitable approximation of the
low-energy dynamics of SYM or not thus depends on the details of the full
theory. Of main importance is the question whether $\Delta \mathcal{L}$ really
contains derivative terms on the field $F$ or not.
Such terms must be present due to the following two points:
\begin{enumerate}
\item \label{enum:p1}Supersymmetry is realized linearly. This is one of the basic assumptions
  made when writing down the Lagrangian \eqref{eq:vylagrangian}. But again,
  this is a statement about the whole effective action. Thus not solely
  \eqref{eq:vylagrangian}, but also $\Delta \mathcal{L}$ must be an integral
  over superspace.

  If the low energy dynamics are
  described by $\Phi$ alone (e.g.\ in a quantum effective action), higher
  derivatives on $\varphi$ and $\psi$ inevitably lead to derivative terms in $F$. When introducing
  additional fields, we could try to arrange them in such a way that all
  derivative terms in the auxiliary fields cancel (such a model has been
  derived in ref.\ \cite{gates97} as an ansatz for the effective action of
  SUSY QCD). Using this idea in pure SYM would imply that the new fields
  are not suppressed compared to $\Phi$, in contradiction to our assumption
  about the low-energy dynamics of SYM. Thus it is impossible to keep $F$
  auxiliary in a non-local Lagrangian, if the low-energy dynamics shall depend
  on the three fields $\varphi$, $\psi$ and $F$, only. 
\item \label{enum:p2}The effective action must have higher order derivative terms that are
  not present in the Veneziano-Yankielowicz Lagrangian. A local effective
  action is not acceptable for an interacting theory.

  Besides this general objection a careful analysis within the supersymmetric
  framework shows that the locality of supersymmetric non-linear sigma models
  is not just a harmless peculiarity when using them in effective
  descriptions. Indeed, the low energy spectrum should include the glue-ball. The only candidate for a
  glue-ball operator is $F_{\mu \nu} F^{\mu \nu}$, the real part of $F$. If
  $F$ remains auxiliary we have to assume that the glue-ball of $N=1$ SYM does not
  belong to the relevant low-energy degrees of freedom. Even if this
  assumption is correct the relation to pure YM theory remains
  mysterious. If the low-energy dynamics shall be described in terms of
  renormalizable operators (or their classical fields, resp.)\ pure YM theory
  does generate non-trivial dynamics for $F_{\mu \nu}
  F^{\mu \nu}$ and we are left with the unsatisfactory situation that either
  the two classical fields $\bra{\Omega(J)} F_{\mu \nu}
  F^{\mu \nu} \ket{\Omega(J)}_{\mbox{\tiny{SYM}}}$ and $\bra{\Omega(J)} F_{\mu \nu}
  F^{\mu \nu} \ket{\Omega(J)}_{\mbox{\tiny{YM}}}$ are completely independent objects
  or that
    there exists a phase transition in the decoupling of the gluino by giving
    it a heavy mass (its origin and consequences have been discussed in
    \cite{bergamin01}). It has been tried to escape this restriction by the
    choice of a more complicated geometry than realized in
    \eqref{eq:vylagrangian} \cite{farrar98}. From our point of view the result
    of \cite{farrar98} cannot lead to a resolution of the problem
    \cite{bergamin01}. Moreover, trying to add higher order derivatives to the
  action of \cite{farrar98} leads to the same complications as discussed in
  this work for the Veneziano-Yankielowicz Lagrangian and their resolution  automatically leads to the inclusion of the glue-ball dynamics (see
  next section). Thus the construction of \cite{farrar98} becomes redundant in
  the context of non-local actions.
\end{enumerate}
As a consequence of (1) and (2) $F$ has to be seen as an independent degree of freedom. At least in the
framework of effective actions it is possible to formulate consistent
supersymmetric actions of this type \cite{bergamin02:1}, where some important features
of supersymmetric theories with ``normal'' auxiliary fields are compromised. This especially
includes (see ref.\ \cite{bergamin02:1} for a detailed discussion):

\begin{enumerate}
\item The full potential is not positive semi-definite: Obviously the
  positivity of the potential holds after elimination of the auxiliary fields,
  only. The statement that supersymmetric potentials are always positive
  semi-definite is actually a statement about the maximum of the auxiliary
  field potential: The latter is zero (unbroken supersymmetry) or positive
  (broken supersymmetry). As this maximum does no longer represent the ground
  state in a theory with dynamical ``auxiliary'' fields, the physical potential
  can have a negative minimum. This has important consequences in the
  discussion of supersymmetry breaking: In classical and perturbative
  supersymmetry a supersymmetric state (i.e. a state with vanishing potential) is automatically the ground-state and
  implies unbroken supersymmetry. In the non-perturbative region this
  conclusion does not hold, as the ``auxiliary'' fields become dynamical.
\item While the potential is not necessarily positive, the restriction on the
  vacuum energy still holds: $\lomega T_{\mu \nu} \romega = g_{\mu \nu} E_0
  \geq 0$ due to supersymmetry current-algebra relations \cite{salam74}. $E_0
  > 0$ implies supersymmetry breaking, while $E_0 = 0$ means
  unbroken supersymmetry. In a strictly perturbative and/or semiclassical
  logic a strict relation between the minimum of the effective potential and
  the order parameter $E_0$ exists: $(\veff)_0 \equiv E_0$ due to the
  non-renormalization theorem. In a fully non-perturbative framework an
  extension of this relation to $(\veff)_0 \neq 0$, $E_0 \neq 0$ but
  $(\veff)_0 \neq E_0$ is possible. A consistent relation of this form can
  be achieved only by a non-perturbative extension of the trace anomaly
  \eqref{eq:anomcurrent}\footnote{In this paper we do not provide such an extension.}.
  But in any case the order parameter of
  supersymmetry breaking is still $E_0$ and in the effective theory the
  latter is represented by the value of
  $F$ in the ground-state.
\item Supersymmetry breaks dynamically. Interestingly enough this follows directly
  from \eqref{eq:vylagrangian} and \eqref{eq:vyonshell}: The unstable maximum
  is at the supersymmetry preserving point (vanishing potential and $F = 0$). As the
  action \eqref{eq:vylagrangian} is complete up to order
  $\mathcal{O}(\Lambda^{-4})$, the unstable maximum cannot turn directly into
  a minimum (stability of $\mathcal{O}(p^2)$ forbids to choose $a<0$ in \eqref{eq:vylagrangian}). Instead, the new minimum must be at a different value of $F$:
  $F_0 \neq 0$. But $F$ is the order parameter of supersymmetry breaking and
  we arrive at the desired result. The argument of \cite{veneziano82} that
  supersymmetry is unbroken in $N=1$ SYM thus turns into the opposite when
  taking into account all (including non-local) terms.
\end{enumerate}

%%% Local Variables: 
%%% mode: latex
%%% TeX-master: "main"
%%% End: 

%% file: construction.tex
\section{Non-Local Effective Action of SYM}
\label{sec:construction}
As outlined above the extension of the Veneziano-Yankielowicz Lagrangian to a
non-local model with dynamical ``auxiliary'' field changes its fundamental
properties. Thus the explicit form of such an extension must be studied, even
if most of the terms will be unimportant in any application. To this end, a
more rigorous definition of the object discussed here should be given first.
\subsection{Definition and Symmetries of the Action}
\label{sec:constructionone}
In this work we consider a first step in the discussion of non-perturbative SYM
theories. It consists in finding the correct ground-state. The latter is
defined as a thermodynamical limiting process of the
quantum effective action leading to the minimum of the effective potential
(see sect.\ 2 of ref.\ \cite{bergamin01} and references therein). Thus a quantum effective action
must be defined, where the components of $\Phi$ represent the classical
fields. A quantum effective action with this field content is obtained by extending the complex coupling
constant $\tau$ to a chiral superfield of sources \cite{bib:mamink,bib:markus}
\begin{equation}
\label{eq:sourcemult}
J(x) = \tau(x) + \theta \eta(x) - 2
\theta^2 m(x)\ .
\end{equation}
The effective action is then defined as the Legendre transform with respect
to these sources
\begin{equation}
  \label{eq:effactdef}
    \Gamma[\tilde{J},\Tilde{\Bar{J}}] = \intd{\diff{^4x}} \bigl( J(x) \varfrac[W{[}J{]}]{J(x)} +
  \hc \bigr) - W[J,\bar{J}]\ .
\end{equation}
$\tilde{J}$ are the classical fields $\varphi_{\mbox{\tiny cl}}$,
$\psi_{\mbox{\tiny cl}}$ and $F_{\mbox{\tiny cl}}$. As they shall transform linearly under
supersymmetry they can be combined to the chiral multiplet $\Phi$ of equation
\eqref{eq:Phidef} (we suppress the index ``cl'' in the following). \eqref{eq:effactdef} together with \eqref{eq:sourcemult}
defines the unique effective action in terms of gauge-invariant and
supersymmetry covariant classical fields. In the thermodynamical limiting
process
\begin{align}
\label{eq:thdyn}
  \varfrac{\tilde{J}} \Gamma[\tilde{J},\Tilde{\Bar{J}}] &= J(x) \rightarrow 0 & \varfrac{J(x)} W[J,\bar{J}]
  &= \tilde{J}(x) \rightarrow \tilde{J}\cc
\end{align}
this effective action (or its effective potential resp.)\ lead per
definitionem to the true ground-state of the theory. Equation \eqref{eq:thdyn}
has to be understood as a limiting process starting from a suitable UV and IR
regularization and performing the full non-perturbative
renormalization. During this process \emph{all} regularizations are getting
removed and the sources are finally relaxed to $J(x) \rightarrow 0$. The dual
parameters (classical fields) adopt their correct vacuum expectation values in
the limit.

Of course, we cannot perform these calculations explicitly. Thus certain assumptions enter in writing down equation
\eqref{eq:effactdef} as discussed in
\cite{bergamin01} and references therein.

Our task is therefore to extend the action \eqref{eq:vylagrangian} to a
non-local model, taking into account the important new features appearing
therein according to the discussion of the last section. This non-local action
is then seen as an ansatz for the quantum effective action
\eqref{eq:effactdef}.

To extend the model \eqref{eq:vylagrangian} to a non-local Lagrangian we use
as starting-point the result of
\cite{bergamin02:1}. Therein the authors showed that the action
\eqref{eq:localkahler} is --within the context of effective field theories-- a
restricted form of a more general model, which is mandatory and of the form
\begin{equation}
  \label{eq:nonlocalkahler}
    \mathcal{L} = \intd{\diff{^4x}} \biggl( \intd{\diff{^4 \theta}} K(\Psi_n,
  \bar{\Psi}_n) - \bigl(\intd{\diff{^2 \theta}} W(\Psi_n) + \hc \bigr) \biggr)\ .
\end{equation}
The index $n$ runs from zero to infinity, the defining field is $\Psi_0 \equiv
\Psi$ and the higher $\Psi_n$ are related to $\Psi_0$ by
\begin{align}
  \label{eq:psincondition}
  \Psi_n &= \bar{D}^2 \bar{\Psi}_{n-1}\ , & \Psi_{2n} &= (-1)^n \Box^n \Psi_0\ , & \Psi_{2n +1}
  &= (-1)^n \Box^{n} \Psi_1\ .
\end{align}
Of course, the Lagrangian in \eqref{eq:nonlocalkahler} could be formulated in terms of $\Psi_0$ and $\Psi_1$ alone  when
dropping the assumption that the K\"{a}hler- and superpotential are
polynomials in $\Psi$, but are allowed to include explicit space-time
derivative terms. The corresponding action \eqref{eq:intro1}
\begin{equation}
  \label{eq:nlnonkahler}
  \mathcal{L} = \intd{\diff{^4x}} \biggl( \intd{\diff{^4 \theta}} A(\Psi_0,\Psi_1 , \bar{\Psi}_0, \bar{\Psi}_1) - \bigl(
  \intd{\diff{^2 \theta}} H(\Psi_0) + \hc \bigr) \biggr)
\end{equation}
is actually more general than \eqref{eq:nonlocalkahler}. However, the effective
potential from \eqref{eq:nonlocalkahler} and \eqref{eq:nlnonkahler} are the
same, which is the quantity we are mostly interested in. The advantage of
\eqref{eq:nonlocalkahler} is its simple mathematical form, but in certain
cases stability requirements
force us to add one term of order $\mathcal{O}(p^2)$ that transgresses this
form (see section \ref{sec:consdyn} below).

The defining superfield of our model is the classical field $\Phi$ obtained from
Legendre transformation. Nevertheless it is difficult to implement all
symmetries when choosing $\Psi = \Phi$. Standard superconformal transformations of a chiral
superfield imply that the mass dimension and
the $R$ weight of such a  field obey the relation $d = (3/2) r$. Of course,
$\Phi$ is a correct chiral superfield in this sense, it has mass dimension 3
and chiral weight 2. But $\bar{D}^2 \bar{\Phi}$ has mass dimension 4 and
chiral weight 0. The field still has standard transformations under $R$
symmetry and scale transformations, but not under S-supersymmetry. Thus an
action in terms of $\Phi$ and $\bar{D}^2 \bar{\Phi}$ invariant under $R$ and
scale transformations would not be invariant under S-supersymmetry.

We can escape these difficulties by observing that a chiral field $C$ of the
form $C = \bar{D}^2 \bar{B}$ transforms under S-supersymmetry with
dimension 2, regardless of the actual mass dimension of $C$. If the true
dimension of $C$ is not equal to 2, there appear additional anomalous
transformation terms. Thus we recover standard transformations under the full
superconformal group by choosing
\begin{align}
\label{eq:thirdroot}
  \Psi_0 &= (\Phi)^{\frac{1}{3}}\ .
\end{align}
A detailed analysis of the superconformal Ward identities, leading to the
defining superfields $\Psi_0$ and $\Psi_1$ as given in \eqref{eq:thirdroot},
has been performed in \cite{Shore:1983kh}.

Of course, the third root of equation \eqref{eq:thirdroot} has a threefold
multivaluedness. The consequential ambiguities in the invariant part of the
Lagrangian are the same as in \eqref{eq:vylagrangian}. Due to the relation
\eqref{eq:thirdroot} the WZ-term of the actions \eqref{eq:nonlocalkahler} and
\eqref{eq:nlnonkahler} is restricted to the result found by Veneziano
and Yankielowicz, which does not show an ambiguity.

K\"{a}hler and superpotential in \eqref{eq:nonlocalkahler} are simply
polynomials in the fields $\Psi_n$ and thus we may write the invariant part of this
action as
\begin{align}
\label{eq:kahlerexp}
  K(\Psi_n, \bar{\Psi}_n) &= \sum_{\xi \bar{\xi}} \alpha_{\xi \bar{\xi}}(\Psi_0)^{\xi_0} \cdots
  (\Psi_\infty)^{\xi_\infty} (\bar{\Psi}_0)^{\bar{\xi}_0} \cdots
  (\bar{\Psi}_\infty)^{\bar{\xi}_\infty}\ , \medsp
\label{eq:suppotexp}
  W(\Psi_n) &= \sum_{\chi} \beta_{\chi} (\Psi_0)^{\chi_0} \cdots
  (\Psi_\infty)^{\chi_\infty}\ .
\end{align}
The theory depends on infinitely many dimensionless coupling constants
$\alpha_{\xi \bar{\xi}}$ and $\beta_{\chi}$. The vectors $\xi$, $\bar{\xi}$ and $\chi$ are restricted to combinations
with correct mass dimension and $R$ invariance:
\begin{align}
\label{eq:kahlercond}
  \sum_i (i+1) (\xi_i + \bar{\xi}_i) &= 2 & \sum_i \bigl( (\xi_{2i} -
  \bar{\xi}_{2i}) + 2 (\xi_{2i+1} -
  \bar{\xi}_{2i+1})\bigr) &= 0 \medsp
\label{eq:suppotcond}
  \sum_i (i+1) \chi_i &= 3 & \sum_i \bigl( \chi_{2i} + 2 \chi_{2i+1}\bigr) &= 3
\end{align}
In principle each term of the action can now be calculated from the
$\theta$-expansion of $\Psi_0$ and $\Psi_1$ by using the recursive formulas \eqref{eq:psincondition}. We denote the components of
$\Psi_0$ by a hat. In terms of the defining fields $\varphi$,
$\psi$ and $F$ these fields are given by
\begin{align}
\label{eq:psi0exp}
\begin{split}
  \Psi_0 &= \hat{\varphi} + \theta \hat{\psi} + \theta^2 \hat{F} \medsp
         &= \varphi^{\frac{1}{3}} + \inv{3} \varphi^{- \frac{2}{3}} \theta
  \psi +  \inv{3} \theta^2 \bigl( \varphi^{- \frac{2}{3}} F + \inv{3}
  \varphi^{- \frac{5}{3}} (\psi \psi)\bigr)
\end{split} \medsp
\label{eq:psi1exp}
\begin{split}
  \Psi_1 &= \Hat{\Bar{F}} - i \theta \sigma^\mu \partial_\mu \Hat{\Bar{\psi}}
  - \theta^2 \Box \Hat{\Bar{\varphi}} \medsp
  &= \frac{1}{3} \bigl(  \bar{\varphi}^{- \frac{2}{3}} \bar{F} + \frac{1}{3}
  \bar{\varphi}^{- \frac{5}{3}} (\bar{\psi} \bar{\psi})\bigr) - \frac{i}{3} \theta
  \sigma^\mu \partial_\mu\bigl(\bar{\varphi}^{- \frac{2}{3}}\bar{\psi}\bigr) -
  \theta^2 \Box \bar{\varphi}^{\frac{1}{3}}
\end{split}
\end{align}

A similar closed form for the action \eqref{eq:nlnonkahler} cannot be given,
but the explicit derivation of some parts thereof is straightforward on the
same lines. \eqref{eq:kahlerexp} and \eqref{eq:suppotexp} together with
\eqref{eq:kahlercond} and \eqref{eq:suppotcond} define the invariant part of
the effective action, only. Adding a term generating the correct anomalies terminates the construction of
the effective action.
\subsection{The Effective Potential}
In this section we concentrate on the effective potential and derive different
minimum and consistency conditions thereof. In section \eqref{sec:consdyn} below we
prove that stable dynamics around any of its consistent minima can be
defined.

As all $\Psi_n$ with $n>1$ vanish for static configurations the relevant
$\xi$, $\bar{\xi}$ and $\chi$ of eqs.\ \eqref{eq:kahlerexp} and
\eqref{eq:suppotexp} are simply two-vectors. Implementing the constraints
\eqref{eq:kahlercond}/\eqref{eq:suppotcond}, the invariant contributions
from K\"{a}hler and superpotential to the effective potential read:
\begin{align}
\label{eq:statickahlerexp}
  \bigl. K(\Psi_l, \bar{\Psi}_l) \bigr|_{p = 0} &= \sum_{m,n}\alpha_{mn} \Psi_0^{1-2m}
  \Psi_1^m\; \bar{\Psi}_0^{1-2n} \bar{\Psi}_1^n \medsp
\label{eq:staticsuppotexp}
  \bigl. W(\Psi_l) \bigr|_{p = 0} &= \sum_k \beta_k \Psi_0^{3 - 2k} \Psi_1^k
\end{align}
Integrating out superspace yields the effective potential in terms of the
components of $\Psi$ \cite{bergamin02:1}:
\begin{equation}
  \label{eq:effpot1}
  \begin{split}
    \veff &= - \metrhat \hat{F} \Hat{\Bar{F}} + \half{1}
    \metrhat[, \Hat{\Bar{\varphi}}] \hat{F} (\Hat{\Bar{\psi}} \Hat{\Bar{\psi}}) + \half{1}
    \metrhat[, \Hat{\varphi}] \Hat{\Bar{F}} (\Hat{\psi} \Hat{\psi}) -
    \inv{4} \metrhat[, \hat{\varphi} \Hat{\Bar{\varphi}}] (\Hat{\psi}
    \Hat{\psi}) (\Hat{\Bar{\psi}} \Hat{\Bar{\psi}}) \medsp
    &\quad  + \hat{F} W_{, \hat{\varphi}}
    + \Hat{\Bar{F}} \bar{W}_{, \Hat{\Bar{\varphi}}} - \half{1}  (\Hat{\psi}
    \Hat{\psi}) W_{, \hat{\varphi} \hat{\varphi}} - \half{1}  (\Hat{\Bar{\psi}}
    \Hat{\Bar{\psi}}) \bar{W}_{, \Hat{\Bar{\varphi}} \Hat{\Bar{\varphi}}}
  \end{split}
\end{equation}
We use the notation of \cite{bergamin02:1}, where $\hat{\varphi}$ refers to
derivatives with respect to $\Psi_0$ and $\Hat{\Bar{F}}$ to derivatives of
$\Psi_1$. With \eqref{eq:statickahlerexp} and \eqref{eq:staticsuppotexp} the
explicit expression for \eqref{eq:effpot1} in terms of the defining fields
becomes\footnote{The ambiguities showing up in the following equation deserve
  a special discussion --here, in QCD as well as in the Veneziano-Yankielowicz
  Lagrangian-- deferred for the time being.}:
\begin{equation}
  \label{eq:effpot2}
  \begin{split}
    \veff &= - \sum_{m,n} \tilde{\alpha}_{mn} \cdot (\bar{\varphi} \varphi)^{-
    \frac{2}{3} (m+n+1)} \bar{F}^{m+1} F^{n+1} \cdot \medsp
  &\quad \phantom{ - \sum_{m,n} \tilde{\alpha}_{m,n}} \cdot \biggl( 1 + \frac{m+n+1}{3}
    \bigl( (\bar{F}\bar{\varphi})^{-1} (\bar{\psi} \bar{\psi}) +
    (F\varphi)^{-1} (\psi\psi)\bigr)\medsp
    &\quad \phantom{ - \sum_{m,n} \tilde{\alpha}_{m,n}\cdot \biggl(}  + \frac{(m+n+1)^2}{9} (\bar{F}
    F)^{-1} (\bar{\varphi} \varphi)^{-1} (\psi \psi) (\bar{\psi} \bar{\psi})
    \biggr) \medsp
    &\quad + \sum_k \tilde{\beta}_k \cdot (\bar{\varphi} \varphi)^{-
      \frac{2}{3} k} F \bar{F}^k \cdot \medsp
    &\quad \phantom{+ \sum_k \tilde{\beta}_k} \cdot \biggr( 1 + \frac{k}{3} \bigl( (\bar{F}\bar{\varphi})^{-1} (\bar{\psi} \bar{\psi}) +
    (F\varphi)^{-1} (\psi\psi)\bigr) + \frac{k^2}{9} (\bar{F}
    F)^{-1} (\bar{\varphi} \varphi)^{-1} (\psi \psi) (\bar{\psi} \bar{\psi})
    \biggr) \medsp
 &\quad + \sum_l \Tilde{\Bar{\beta}}_l \cdot (\bar{\varphi} \varphi)^{-
      \frac{2}{3} l} F^l \bar{F} \cdot \medsp
    &\quad \phantom{+ \sum_k \Tilde{\Bar{\beta}}_l} \cdot \biggr( 1 + \frac{l}{3} \bigl( (\bar{F}\bar{\varphi})^{-1} (\bar{\psi} \bar{\psi}) +
    (F\varphi)^{-1} (\psi\psi)\bigr) + \frac{l^2}{9} (\bar{F}
    F)^{-1} (\bar{\varphi} \varphi)^{-1} (\psi \psi) (\bar{\psi} \bar{\psi})
    \biggr)   
  \end{split}
\end{equation}
Hermiticity requires $(\tilde{\alpha}_{mn})\dega = \tilde{\alpha}_{nm}$ and
$(\tilde{\beta}_l)\dega = \Tilde{\Bar{\beta}}_l$. Also we have absorbed some
numerical factors in the dimensionless coupling constants. The relations to the
quantities appearing in eqs.\ \eqref{eq:statickahlerexp} and
\eqref{eq:staticsuppotexp} are:
\begin{align}
  \tilde{\alpha}_{mn} &= \frac{(2m - 1)(2n - 1)}{3^{m+n+2}} \alpha_{mn} &
  \tilde{\beta}_k &= \inv{3^{k+1}} \beta_k
\end{align}

Of course the effective potential \eqref{eq:effpot2} is subject to additional
constraints to be discussed below. Nevertheless, we want to make some comments
on this most general form:
\begin{itemize}
\item Already eq.\ \eqref{eq:effpot2} shows that the complete
  Veneziano-Yankielowicz Lagrangian \eqref{eq:vylagrangian} is part of the generalized non-local
  version as well. Indeed, the K\"{a}hler potential and the invariant part of
  the superpotential are given by the terms $\propto \alpha_{00}$ and $\propto
  \beta_0$.
\item Most terms from the superpotential \eqref{eq:staticsuppotexp} are redundant as they appear in the
  K\"{a}hler potential already. This is not surprising as a superpotential
  $\propto (\Psi_1)^n$ can be transformed into a full superspace integral
  $\propto \Psi_0 (\Psi_1)^{n-1}$, which is now a contribution to the K\"{a}hler
  potential. Therefore we will set the invariant part of the superpotential to
  zero in the following, except for the linear term $\propto \beta_0$. To
  avoid linear contributions from the WZ-term $\tilde{\beta}_0 = -c = - 2 \beta /(9 \cg g^3)$
  must be chosen. 
\end{itemize}

To abbreviate the remaining invariant effective potential we define a new
metric in the fundamental fields $\varphi$ and $F$:
\begin{equation}
  \label{eq:tildemetric}
  \metrtilde = \sum_{m,n} \tilde{\alpha}_{mn} (\bar{\varphi}
  \varphi)^{-\frac{2}{3}(m+n+1)} F^n \bar{F}^m = \inv{9} (\bar{\varphi}
  \varphi)^{-\frac{2}{3}} \metrhat
\end{equation}
Then the invariant effective potential again takes the simple form as given in
\eqref{eq:effpot1}. Finally, adding the WZ-term the complete effective
potential becomes
\begin{equation}
  \label{eq:effpot3}
  \begin{split}
    \veff &= - \metrtilde F \Bar{F} + \half{1}
    \metrtilde[, \bar{\varphi}] F (\Bar{\psi} \Bar{\psi}) + \half{1}
    \metrtilde[, \varphi] \Bar{F} (\psi \psi) -
    \inv{4} \metrtilde[, \varphi \Bar{\varphi}] (\psi
    \psi) (\Bar{\psi} \Bar{\psi}) \medsp
    &\quad + c \bigl( F \log \frac{z \varphi}{\Lambda^3} + \bar{F} \log \frac{\bar{z} \bar{\varphi}}{\Lambda^3}
      - \inv{2 \varphi} (\psi \psi)  - \inv{2 \bar{\varphi}} (\bar{\psi}
    \bar{\psi})  \bigr)\ .
  \end{split}
\end{equation}
Choosing $\metrtilde = (\bar{\varphi} \varphi)^{-\frac{2}{3}}$,
\eqref{eq:effpot3} reduces exactly to the potential of
\eqref{eq:vylagrangian}. But in contrast to  \eqref{eq:vylagrangian}
$\metrtilde$ is now a function of $\varphi$ and $F$, where any --positive or
negative-- power in $F$ may appear. Therefore it is possible to obtain extrema
of \eqref{eq:effpot3} that are minima in both, $\varphi$ and $F$. 

Starting from equation \eqref{eq:effpot3} these extrema and the corresponding stability conditions
are equivalent to those derived in \cite{bergamin02:1}, when replacing the
original metric $\metrhat$ by the new one $\metrtilde$. The minima in the field
$F$ are found for
\begin{equation}
\label{eq:Fpotconstr}
\begin{split}
  \Bigl. \metrtilde \bar{F} + \metrtilde[, F] F \bar{F}\Bigr|_{F = F_0} &= c
  \log \frac{z \varphi}{\Lambda^3}\ , \medsp
  \Bigl. \metrtilde + \metrtilde[, F \bar{F}] F \bar{F} + \bigl( \metrtilde[, F] F +
  \hc \bigr)\Bigr|_{F = F_0} &< 0\ .
\end{split}
\end{equation}
The complex number $z$ must again be chosen according to the discussion in
section \ref{sec:vylag}. If $F_0 \neq 0$ we may replace the second constraint by
\begin{equation}
  \label{eq:Fpotconstr2}
  \Bigl. \metrtilde - \metrtilde[, F \bar{F}] F \bar{F} - c \bigl(F^{-1} \log \frac{z \varphi}{\Lambda^3} + \hc
  \bigr) \Bigr|_{F = F_0} > 0\ .
\end{equation}
The analogue conditions for $\varphi$ are
\begin{align}
\label{eq:phipotconstr}
  \Bigl. \bar{F} \varphi \metrtilde[, \varphi] \Bigr|_{\varphi = \varphi_0} &=
  c\ , & \Bigl.  F
  \bar{F} \metrtilde[, \varphi \bar{\varphi}] \Bigr|_{\varphi = \varphi_0} &<
  0\ .
\end{align}
The constraint $\metrtilde[, \varphi \bar{\varphi}] < 0$ follows from the
stability of the four-Fermi interactions as well. The mass of $\psi$ --the Goldstino--
vanishes due to the first constraint in \eqref{eq:phipotconstr}. This is an
alternative proof of dynamical supersymmetry breaking. A consistent
ground-state must obey  the constraints \eqref{eq:Fpotconstr} and
\eqref{eq:phipotconstr}. This is possible if and only if there exists a
massless spinor $\psi$. But this must be a Goldstone mode if the theory is not
free (even with a ``supersymmetric'' spectrum with both, $\psi$ and $\varphi$
massless \cite{bergamin02:1}). There is an unique consequence: supersymmetry breaks
dynamically. Note that the above steps do not assume a certain value for
$F_0$. $F_0 \neq 0$ can be seen as a consequence of the condition \eqref{eq:Fpotconstr2} and
the dynamics of $\psi$.

Equations \eqref{eq:Fpotconstr}-\eqref{eq:phipotconstr} illustrate the
essential change of the
role of holomorphic and non-holomorphic parts of the effective
potential. Indeed, a consistent spectrum (massive spectra for $\varphi$ and $F$) is
possible even for $c = 0$ (i.e.\ vanishing superpotential). In contrast to the
Veneziano-Yanlielowicz model (sect.\ \ref{sec:vylag}) the superpotential
reduces to the generator of the anomalies but \emph{never} defines stable
couplings in all fields. The stabilization of the potential for large values
of the fields essentially includes non-holomorphic couplings. As we cannot
eliminate the ``auxiliary'' fields --which can now be seen as a consequence of the
first equation in \eqref{eq:Fpotconstr}-- these couplings must stem from the
K\"{a}hler potential alone.

Finally we should implement two constraints from the underlying
dynamics. First, current algebra relations of supersymmetry tell us that $F_0
\geq 0$ and second $\varphi_0 < 0$ due to vacuum alignment. Obviously it is
impossible to reduce these two conditions to simple conditions on the coupling
constants $\tilde{\alpha}_{mn}$. An appealing (though not necessary) condition
is to set all off-diagonal terms of $\alpha_{mn}$ to zero:
\begin{align}
  \alpha_{mn} &= 0 \ \ (m\neq n)\ , & z&= -1 \ .
\end{align}
The potential then reduces to
\begin{multline}
  \label{eq:effpot4}
    \veff = - \sum_{m} \tilde{\alpha}_{mm} \cdot (\bar{\varphi} \varphi)^{-
    \frac{2}{3} (2m+1)} (\bar{F} F)^{m+1} \cdot \medsp
  \cdot \biggl( 1 + \frac{2m+1}{3}
    \bigl( (\bar{F}\bar{\varphi})^{-1} (\bar{\psi} \bar{\psi}) +
    (F\varphi)^{-1} (\psi\psi)\bigr)  + \frac{(2m+1)^2}{9} (\bar{F}
    F)^{-1} (\bar{\varphi} \varphi)^{-1} (\psi \psi) (\bar{\psi} \bar{\psi})
    \biggr) \medsp
    + c \bigl( F \log\bigl( -  \frac{\varphi}{\Lambda^3}\bigr) + \bar{F} \log
    \bigl(- \frac{ \bar{\varphi}}{\Lambda^3}\bigr)
      - \inv{2 \varphi} (\psi \psi)  - \inv{2 \bar{\varphi}} (\bar{\psi} \bar{\psi})  \bigr)
\end{multline}
and the remaining coupling constants still have to be arranged in such a way
that a minimum with $F_0 > 0$ results.

The principle steps of this section leading to the effective potential
\eqref{eq:effpot3} had been performed in \cite{Shore:1983kh} already. However,
the author did not derive the minimum conditions
\eqref{eq:Fpotconstr}-\eqref{eq:phipotconstr}, but expected a spectrum with
unbroken supersymmetry ($F_0 = 0$). Excluding inverse powers of $F$ he
concluded that the model must be unstable and he did not find the correct stable
ground-state. This agrees with our analysis of
\cite{bergamin02:1} and of the present work. Indeed, the description breaks
down as $F \rightarrow 0$.
\subsection{Consistent Dynamics}
\label{sec:consdyn}
In the last section we discussed the properties of the non-perturbative
effective potential, where $F$ becomes a dynamical variable. However, we have not yet shown that these
dynamics can be introduced consistently when starting from an effective
potential of the form \eqref{eq:effpot3}. In this section we now prove that
this is possible for any ground-state consistent with the constraints derived
in the last section.  We cannot fix all coefficients
in a momentum-expansion around some assumed ground-state as well as we are not able to determine the correct
choice of the coupling constants $\alpha_{mn}$. However, this is not
of main importance: While we do know that some choice of the $\alpha_{mn}$
leads to the correct ground-state, the effective action may not include the
whole physical dynamics around this state. But we do know that there are
some dynamics and that they must be stable (see ref.\ \cite{bergamin01} for a
detailed discussion).

In reference \cite{bergamin02:1} it has been shown that acceptable dynamics of
this type are possible if
the defining fields are the fields appearing in the action
\eqref{eq:nonlocalkahler} or \eqref{eq:nlnonkahler}. In other words, the
existence of stable dynamics is known on the level of the fields
$\hat{\varphi}$, $\hat{\psi}$ and $\hat{F}$. The strategy is as follows: First
the dynamics of the action from \eqref{eq:statickahlerexp} are
calculated. That means, we determine the effects of non-static field
configurations in the K\"{a}hler potential \eqref{eq:statickahlerexp}. These
dynamics are expressed in terms of geometric quantities of the K\"{a}hler
potential $K(\hat{\varphi}, \Hat{\Bar{F}}, \Hat{\Bar{\varphi}}, \hat{F})$. Due
  to the essential role of the field $F$ a part of the resulting dynamics can be
  unstable in the minimum of the effective potential. Thus we add additional
  terms to the action, all of them including explicit space-time
  derivatives. Obviously they do not change the effective potential, but they
  allow to add derivative terms of the order $\mathcal{O}(p^n)$ $(n\geq 2)$
  and to remove all instabilities. This procedure is possible as
  $F_0 \neq 0$.

In this way we are able to prove that any effective potential of the form
\eqref{eq:effpot3} with $F_0 \neq 0$ can be provided with stable dynamics. Nevertheless, the
concrete expressions will look rather undetermined in general. Thus we should point
out that \eqref{eq:effpot3} does not fix all coupling constants. Indeed, 
different actions should be considered as
equivalent if they have the following properties in common:
\begin{itemize}
\item The value of the effective potential in the minimum is the same.
\item The vacuum expectation values of the fields $\varphi$ and $F$ are
  the same.
\item In a systematic expansion around the ground-state, the dominant
  couplings (including the spectrum) are the same.
\end{itemize}

In a first step we derive the momentum-expansion of the action
\eqref{eq:nonlocalkahler} with K\"{a}hler and superpotential as given in
\eqref{eq:statickahlerexp} and \eqref{eq:staticsuppotexp}. This expansion
stops at order $\mathcal{O}(p^4)$. The kinetic term for $\hat{\psi}$ is given
by
\begin{equation}
  \label{eq:ord1orig}
  \mathcal{L}^{(1)} = \half{i} \bigl( \metrhat  + \metrhat[, \hat{F}]
    \hat{F}_0 + \metrhat[, \Hat{\Bar{F}}]
    \Hat{\Bar{F}}_0 \bigr)\hat{\psi} \sigma^\mu \lrpartial_\mu
    \Hat{\Bar{\psi}}\ .
\end{equation}
With \eqref{eq:tildemetric} this simply translates into
\begin{equation}
  \label{eq:ord1new}
  \mathcal{L}^{(1)} = \half{i} \bigl(\metrtilde  + \metrtilde[, F] F_0 +
    \metrtilde[, \bar{F}] \bar{F}_0 \bigr) \psi \sigma^\mu \lrpartial_\mu
    \bar{\psi}\ .
\end{equation}
The derivatives of $\mathcal{O}(p^2)$ acting on the spinors are rather simple
as well:
\begin{equation}
  \label{eq:ord2fer}
\begin{split}
  \Ltext{fer}^{(2)} &= ( g_{\hat{\varphi} \hat{F}} + \half{1} g_{\hat{\varphi} \hat{F},
    \hat{F}} \hat{F}_0 ) \hat{\psi} \Box \hat{\psi} + ( g_{\Hat{\Bar{\varphi}} \Hat{\Bar{F}}} +
      \half{1} g_{\Hat{\Bar{\varphi}} \Hat{\Bar{F}},
    \Hat{\Bar{F}}} \Hat{\Bar{F}}_0 ) \Hat{\Bar{\psi}} \Box
    \Hat{\Bar{\psi}}\medsp
    &=  \half{1} \bigl( \tilde{g}_{\varphi F} \psi \Box \psi +
    \tilde{g}_{\bar{\varphi} \bar{F}} \bar{\psi} \Box \bar{\psi} \bigr)
\end{split}
\end{equation}
In the second line we introduced the new notation
\begin{equation}
  \label{eq:gphiF}
  \tilde{g}_{\varphi F} = \inv{9} \varphi_0^{-\frac{4}{3}} \bigl( 2 g_{\hat{\varphi} \hat{F}} + g_{\hat{\varphi} \hat{F},
    \hat{F}} \hat{F}_0 \bigr)
\end{equation}
This new component of the ``metric'' $\tilde{g}$ as well as further components
to be defined below are no longer derivatives of a K\"{a}hler potential, but
optimized to economize writing. The bosonic part of $\mathcal{O}(p^2)$ turns
out to be rather complicated. After transformation to the defining fields it
can be written as
\begin{equation}
  \label{eq:ord2bos}
\begin{split}
  \Ltext{sc}^{(2)} &= \Bigl( \metrtilde + \frac{4}{9} \tilde{g}_{F \bar{F}} (\bar{\varphi} \varphi)_0^{-1} (\bar{F}
  F)_0 + \bigl(\metrtilde[, F] F_0 - \frac{2}{3} \tilde{g}_{\varphi
  \bar{F}} (\bar{\varphi}^{-1} \bar{F})_0 + \hc \bigr) \Bigr) \partial_\mu \bar{\varphi} \partial^\mu \varphi \medsp
  &\quad + \tilde{g}_{F \bar{F}} \partial_\mu \bar{F} \partial^\mu F  + \bigl(
  \tilde{g}_{\varphi F} \partial_\mu F \partial^\mu \varphi + \hc \bigr)
  \medsp
  &\quad + \Bigl( \bigl( \tilde{g}_{\varphi \bar{F}} - \frac{2}{3}
  \tilde{g}_{F \bar{F}} (\varphi^{-1} F)_0 \bigr) \partial_\mu \bar{F}
  \partial^\mu \varphi + \hc \Bigr) \medsp
  &\quad + \Bigl( \bigl( \tilde{g}_{\varphi \varphi} - \frac{2}{3} \tilde{g}_{\varphi
  F} (\varphi^{-1} F)_0 \bigr) \partial_\mu \varphi \partial^\mu \varphi + \hc
  \Bigr)\ .
\end{split}
\end{equation}
The new quantities introduced here are:
\begin{align}
  \tilde{g}_{F \bar{F}} &= \inv{9} (\bar{\varphi} \varphi)_0^{-\frac{2}{3}}
  g_{\hat{F} \Hat{\Bar{F}}} \medsp
  \tilde{g}_{\varphi \bar{F}} &= \inv{9} \hat{F}_0 (\bar{\varphi}
  \varphi)_0^{-\frac{2}{3}} g_{\hat{\varphi} \hat{F}, \Hat{\Bar{F}}}\medsp
  \tilde{g}_{\varphi \varphi} &= \inv{9}  \varphi_0^{-\frac{4}{3}} \hat{F}_0
  g_{\hat{\varphi} \hat{\varphi}, \hat{F}}
\end{align}
Finally the terms $\mathcal{O}(p^3)$ and $\mathcal{O}(p^4)$ read:
\begin{align}
\label{eq:ord3}
      \mathcal{L}^{(3)} &= - \half{i} \tilde{g}_{F \bar{F}} \psi \sigma^\mu
    \lrpartial_\mu \Box \bar{\psi}\medsp
\label{eq:ord4}
    \mathcal{L}^{(4)} &= \tilde{g}_{F \bar{F}} \Box \bar{\varphi} \Box \varphi
\end{align}

Assume now that we have found the correct ground-state of $N=1$ SYM from the
effective potential \eqref{eq:effpot3}, which corresponds to a certain choice
of the coupling constants $\tilde{\alpha}_{mn}$ in \eqref{eq:tildemetric}. The
resulting dynamics may have instabilities in momentum space in \eqref{eq:ord2bos} and (less
important) in \eqref{eq:ord4}. In addition, \eqref{eq:ord2bos} has many
off-diagonal terms that we might want to cancel. In a first step we can
use the freedom in the choice of the $\tilde{\alpha}_{mn}$ mentioned above. In
fact, there is one term only that we cannot change in this way: The
coefficient of $\mathcal{L}^{(1)}$ and the corresponding part of $\partial_\mu \bar{\varphi}
\partial^\mu \varphi$ in $\mathcal{L}^{(2)}$ (cf.\ the comments at the end of
this section). All other terms in the momentum-expansion depend on 
derivatives of the K\"{a}hler potential that do not
appear in the effective potential. This follows from the fact that the
effective potential depends on $\metrtilde$ and derivatives thereof,
only. Thus we can expect that most terms in the momentum expansion are not 
fixed  by some choice of the ground-state. Nevertheless, this
freedom may not be sufficient to obtain an acceptable dynamical behavior. Then
we have to add terms involving explicit space-time derivatives according to
the following rule \cite{bergamin02:1}: To change the behavior of the
Lagrangian at order $\mathcal{O}(p^n)$ we add a term with $n$ space-time
derivatives.

Inspecting the actions \eqref{eq:nonlocalkahler} and \eqref{eq:nlnonkahler}
together with the conditions \eqref{eq:kahlerexp} and \eqref{eq:suppotexp} one
finds:
\begin{itemize}
\item The superpotential remains a polynomial function in the fields.
\item The dynamics to order $\mathcal{O}(p^2)$ cannot be changed by the action
  of the type \eqref{eq:nonlocalkahler}, as a term $\propto \Psi_0 \Box \bar{\Psi}_0$
  does not exist.
\end{itemize}
Using the action \eqref{eq:nlnonkahler}, the new Lagrangian to add has the form
\begin{equation}
  \label{eq:addlag}
  \Ltext{add} = \intd{\diff{^4x}} \intd{\diff{^4 \theta}} \sum_{k = 1}^\infty d_k (\Psi_0
  \bar{\Psi}_0)^{-k} (\partial_\mu \Psi_0 \partial^\mu \Box^{k-1}
  \bar{\Psi}_0)\ .
\end{equation}
If $d_1$ can be chosen zero we can write instead of \eqref{eq:addlag}
\begin{equation}
  \label{eq:addlag2}
  \Ltext{add} = - \intd{\diff{^4x}} \intd{\diff{^4 \theta}} \sum_{k =2}^{\infty} d_k (\Psi_0
  \bar{\Psi}_0)^{-k} (\Box \Psi_0 \Box^{k-1}
  \bar{\Psi}_0)\ ,
\end{equation}
which is now part of the action \eqref{eq:nonlocalkahler}.
This leads to the following new terms bilinear in the fields:
\begin{align}
\label{eq:ladd2}
\begin{split}
  \Ltext{add}^{(2)} &= \intd{\diff{^4x}} \frac{d_1}{9} (\bar{\varphi}
  \varphi)_0^{-1} \bigl( (\bar{F} F)_0 (\bar{\varphi} \varphi)_0^{-1}
  \partial_\mu \bar{\varphi} \partial^\mu \varphi + \partial_\mu \bar{F}
  \partial^\mu F \medsp
  &\quad \phantom{\intd{\diff{^4x}}} - (F \varphi^{-1})_0 \partial_\mu \varphi \partial^\mu
  \bar{F} - (\bar{F} \bar{\varphi}^{-1})_0 \partial_\mu \bar{\varphi}
  \partial^\mu F \bigr)
\end{split}\medsp
\label{eq:laddodd}
    \Ltext{add}^{(2k +1)} &= - \intd{\diff{^4x}} \frac{i d_k}{18}
    (\bar{\varphi} \varphi)_0^{- \frac{k+2}{3}} \psi \sigma^\mu \lrpartial_\mu
    \Box^k \bar{\psi}
\end{align}
\begin{align}
    \begin{split}
    \Ltext{add}^{(2k)} &= - \intd{\diff{^4x}} (\bar{\varphi}
  \varphi)_0^{-\frac{k+1}{3}} \Bigl( \bigl( \frac{(k+2)^2}{9} d_k (\bar{\varphi}
  \varphi)_0^{- \frac{4}{3}} (\bar{F} F)_0 - d_{k-1} \bigr) \varphi \Box^k
  \bar{\varphi}  + d_k (\bar{\varphi}
  \varphi)_0^{-\frac{1}{3}} F \Box^k \bar{F} \medsp
  &\quad \phantom{- \intd{\diff{^4x}}} - \frac{k+2}{3} d_k (\bar{\varphi}
  \varphi)_0^{-\frac{1}{3}} \bigl( (F \varphi^{-1})_0  \varphi \Box^k
  \bar{F} + (\bar{F} \bar{\varphi}^{-1})_0  \bar{\varphi}
  \Box^k  F \bigr) \Bigr)
    \end{split}
\label{eq:laddeven}
\end{align}
In equation \eqref{eq:laddodd} the index runs from $1 \cdots \infty$, in
equation \eqref{eq:laddeven} from $2 \cdots \infty$. The proof that we can
obtain stable dynamics is now rather trivial:
\begin{itemize}
\item If the dynamics to order $\mathcal{O}(p^2)$ are not stable we choose
  $d_1 > 0$ in such a way that the eigenvalues of the dynamics are all
  positive. This is always possible as
  \begin{enumerate}
  \item \eqref{eq:ladd2} has no unstable terms. This follows from $F_0 > 0$
    and $\varphi_0 < 0$. On the other hand it adds stable terms to all possible combinations.
  \item $d_1$ is not constrained.
  \end{enumerate}
\item If $\tilde{g}_{F \bar{F}} > 0$ the dynamics to order $\mathcal{O}(p^4)$
  in $\varphi$ have the wrong sign. Also, the term in $d_1$ adds unstable
  terms in all combinations at this order. These terms are suppressed already, if $p^2
  \ll \Lambda^2$. Nevertheless we might want to change this. We thus choose:
  \begin{equation}
    d_2 > \frac{9}{16} (\bar{\varphi} \varphi)_0^{\frac{4}{3}} (\bar{F}
    F)_0^{-1} \bigl( 9 (\bar{\varphi}
    \varphi)_0 \tilde{g}_{F \bar{F}} + d_1 \bigr)
  \end{equation}
  If $\tilde{g}_{F \bar{F}} < 0$ we relax the constraint for $d_2$ to equation
  \eqref{eq:dcond} below.
\item Now the dynamics are stable to $\mathcal{O}(p^4)$, but $d_2$ adds
  unstable terms to order $\mathcal{O}(p^6)$. In general all unstable terms
  are getting removed by the recursive formula
  \begin{equation}
    \label{eq:dcond}
    d_k > \frac{9}{(k+2)^2} (\bar{\varphi} \varphi)_0^{\frac{4}{3}} (\bar{F}
    F)_0^{-1} d_{k-1} \ .
  \end{equation}
  In addition to the suppression by orders of $\Lambda$ the new coupling
  constants can thus be chosen suppressed by a factor $k^{-2}$.
\end{itemize}
We point out again that this procedure is consistent with the ansatz
\eqref{eq:nonlocalkahler} if $d_1$ can be chosen zero. If so, the
dynamics to $\mathcal{O}(p^2)$ must be stable without further modifications.

To cancel unexpected off-diagonal contributions to the kinetic Lagrangian,
more complicated terms including explicit space-time derivatives can be
introduced. A possible set, which allows to cancel any off-diagonal
contribution, has been given in \cite{bergamin02:1}.

There exists one term in the dynamics that remains unchanged by any choice of
the $d_k$: $\mathcal{L}^{(1)}$. Therefore it is a strict condition on the
effective potential that the coefficient of this term does not vanish in the
minimum. But this simply means
\begin{equation}
  \label{eq:L1cond}
\metrtilde  + \metrtilde[, F] F_0 +
    \metrtilde[, \bar{F}] \bar{F}_0 = - \metrtilde + c \bigl(\bar{F}_0^{-1}
    \log\bigl( \frac{z\varphi}{\Lambda^3}\bigr) + \hc \bigr)  = (\veff)_0 \neq 0\ ,
\end{equation}
where we have used \eqref{eq:Fpotconstr}. This condition just says that the
effective potential in the minimum must be non-vanishing: $(\veff)_0 \neq
0$. Nevertheless, the sign is not fixed by stability arguments. The constraint
then says that our models are actually divided in two classes, one having
$(\veff)_0 > 0$, the other one $(\veff)_0 < 0$. Starting from a model of one
class it is impossible to deform the latter into a model of the other class by
a continuous transformation.

%%% Local Variables: 
%%% mode: latex
%%% TeX-master: "main"
%%% End: 

%% file: discussion.tex
\section{The Role of Dynamical Symmetry Breaking}
\label{sec:DSB}
We have shown in the last section how to construct a non-local effective action
for SYM. We have proven that the inclusion of higher order derivatives breaks
supersymmetry dynamically and we finally showed that any consistent minimum of
the effective potential can be equipped with stable dynamics. Nevertheless several questions remain and in this
section we want to discuss some of them, especially:
\begin{enumerate}
\item Different semi-classical and/or perturbative arguments suggest that
  supersymmetry should be unbroken in SYM theories. Why don't these arguments
  exclude our solution?
\item Closely related to the first question is the behavior of the potential
  for $\varphi \rightarrow 0$ and $F \rightarrow 0$. The first limit is
  interesting in the action by Veneziano and Yankielowicz as well: There the
  bosonic potential has a second minimum for $\varphi = 0$, which led to
  speculations about a chirally symmetric minimum
  \cite{kovner97}. Do similar problems appear in our calculation?
\item We have defined our theory as an effective action obtained from a source
  extension. How does the effective action depend on these
  sources? Especially, we can study the limit of constant sources, which is
  formally identical to soft supersymmetry breaking. What happens to the
  Goldstino in this case?
\end{enumerate}
\subsection{Singularities of $\mathbf{\veff}$ and DSB}
\label{sec:veffsing}
We address the second point first. The bosonic potential is
non-singular for $\varphi = 0$ if $m \leq - \half{1}$ in
\eqref{eq:effpot4}. This allows a term $(\Bar{F} F)^{\half{1}}$ to stabilize
the potential for large $F$ and in principle a consistent minimum should be
possible. Nevertheless, the meaning of the limit $\varphi \rightarrow 0$
remains unclear: As in the case of Veneziano and Yankielowicz, the dynamics
as well as the fermionic potential diverge at this point. Also, the potential
is now restricted to a simple form that we do not expect for an effective
theory. Thus this restriction should not play any role in the discussion.

More interesting is the behavior for $F \rightarrow 0$, as this limit can be
important in the interpretation of dynamical supersymmetry breaking. The
bosonic potential is regular in this limit for $m \geq -1$. Again, this should
be sufficient to construct a consistent minimum in the sense of the conditions
\eqref{eq:Fpotconstr} and \eqref{eq:phipotconstr}. Similar to the situation of
$\varphi \rightarrow 0$, $m \geq -1$ is not enough to ensure a regular limit of the
whole potential. Also, the dynamics will diverge in the limit, which
however is true for any model of the type discussed in this
work. Regularity of the whole potential would require $m \geq 0$, a constraint
that cannot be fulfilled.

The typical behavior expected is thus the divergence of the potential in both
limits, $\varphi \rightarrow 0$ and $F \rightarrow 0$. The divergence for
$\varphi \rightarrow 0$ is related to chiral symmetry breaking and thus to the
impossibility of massive spectra in the fermionic sector if $\varphi =
0$. Similarly we can interpret the divergence for $F \rightarrow 0$ with the
the massive spectrum in boson sector. Although the relation between the vacuum
expectation value of $\tr F_{\mu \nu} F^{\mu \nu}$ and the formation of the
mass-gap in this sector is less well understood than the analogue relation in the
fermionic sector, it seems that the infrared divergences cannot be removed if
this vacuum expectation value vanishes \cite{cox80:1,cox80:2}. For further discussions of the
interpretation of $F_0 \neq 0$ within supersymmetry we refer to
\cite{bergamin01}.

Furthermore the divergence of the potential for $F \rightarrow 0$ has important
implications in the interpretation of dynamical supersymmetry breaking (the
first question formulated above). Indeed,
we should be able to explain why semi-classical analysis (instantons,
monopoles etc.)\ as well as topological arguments (Witten index) do not lead
to the correct result. It has been discussed in \cite{bergamin01} (section
6.1) why these arguments cannot exclude dynamical symmetry
breaking. In summary
there exist the following two loopholes:
\begin{enumerate}
\item The supersymmetric state postulated by the Witten index indeed does
  exist, but this state is not the true ground-state. This is
  consistent with all symmetries: Supersymmetry tells us, that the vacuum
  energy must be positive semi-definite. This does not mean that the effective potential
  is positive semi-definite (cf.\ section \ref{sec:vytononlocal},
  sect.\ 6.1 of \cite{bergamin01} and sect.\ 2.2.1 of \cite{bergamin02:1}).

  In this scenario the
  semi-classical analysis is an expansion around the wrong state and fails to
  capture all relevant effects. Any of our non-local models with a regular
  potential for $F \rightarrow 0$ appears as a realization of this
  situation. Trivially it exists if $(\veff)_0 <
  0$, only.
\item The supersymmetric ground-state does not exist at all. In this context it is
  important to notice that all arguments in favor of unbroken supersymmetry
  are brought forward in an infrared regularized region, either in the finite
  volume (instantons and similar, Witten index) or with a momentum cutoff
  (Wilsonian effective action). As these approaches cannot solve the
  fundamental infrared problem of non-Abelian gauge theories, but just remove
  it by brute force, we indeed expect dynamical effects not captured therein. We
  interpret these effects with the formation of the massive spectrum in the
  boson sector and the dynamics of the glue-ball. Our result suggests
  that these dynamical effects are not consistent with unbroken
  supersymmetry. But as any consistent quantum-field theory must solve the
  infrared problem by appropriate dynamical effects, the perturbatively
  supersymmetric ground-state is removed by a singularity and replaced by the
  correct one enforcing dynamical symmetry breaking.
\end{enumerate}
According to the discussion of the limit $F \rightarrow 0$ we expect a
behavior as outlined in point 2. As the supersymmetric state is getting
removed by a singularity, a restriction
on the value of the effective potential in the minimum does not exist.

It has already
been pointed out in \cite{bergamin02:1} that the kinetic term of the Goldstino
has the wrong sign if $(\veff)_0 < 0$. In principle this is not an instability
of the dynamics, but it leads to difficulties in the interpretation of the
action if one would like to couple additional fields. Together with the
singularity in the potential for $F \rightarrow 0$ we thus expect $(\veff)_0 >
0$.

The same coefficient as in $\mathcal{L}^{(1)}$ appears in $\mathcal{L}^{(2)}$
as well. There exists no strict statement that the dynamics to order
$\mathcal{O}(p^2)$ are unstable if $(\veff)_0 < 0$. Nevertheless, $(\veff)_0
> 0$ leads to additional stable contributions in $\mathcal{L}^{(2)}$ and we
can expect stable dynamics even when choosing $d_1 = 0$ in
\eqref{eq:addlag}. This would lead to the gratifying result that the $N=1$ SYM
effective action can be written in the simple form \eqref{eq:nonlocalkahler}.
\subsection{Local Coupling and Hysteresis Effects: Towards an Understanding of
DSB?}
Finally, we want to say a few words about the dependence of the effective
action on its sources and give an outlook on interesting topics to be
considered.

Of course, all sources have been removed by Legendre
transformation. Nevertheless, we can study the limit of (almost) constant
sources, which corresponds to softly broken supersymmetry. In this limit we
formally obtain a parametrical dependence of the effective action on the
constant part of the
sources. This parametrical dependence is not constrained by holomorphic
dependence and the different components of the sources do not form a chiral
superfield (ref.\ \cite{bergamin01}, sect.\ 2 for details). This can be seen as a consequence of renormalization, which does
not respect holomorphic dependence in the local coupling superfield as the
$\beta$-function is essentially non-holomorphic. Notice again
that we are talking about a fully renormalized theory, where the Wilsonian
1-loop $\beta$-function is obsolete.

Especially interesting is the limit of a constant gluino mass $m$ in
\eqref{eq:sourcemult}. Actually the true ground-state of the theory can be
determined by considering hysteresis lines of this type only (ref.\
\cite{bergamin01} and references therein). Thus we should consider all
quantities of our effective action to depend parametrically on a mass
parameter $m$. Of course, the same applies to the Yang-Mills coupling constant
$g$. This extension alone does not necessarily lead to additional constraints on the
effective action. However, by performing the limit $m \rightarrow
\mbox{const.}$ we break supersymmetry softly and thus the (pseudo-)Goldstino should
become massive. This is possible by including a spurion field, only. In
contrast to the spurion field of the fundamental theory, $J = \theta^2 m$, the
spurion field of the effective theory cannot be chiral. This is again a
consequence of the consistent interpretation of the superpotential: In the
context of our effective action, it is
impossible to obtain stable couplings from the superpotential. There exists no
procedure (as e.g.\ the elimination of the auxiliary fields) that would transform a
holomorphic expression like $m \varphi F$ into a non-holomorphic but stable one
($|m|^2 |\varphi|^2$). Obviously, this is not restricted to the fundamental
superpotential but also applies to contributions from spurion fields. Thus, the simplest
possibility would be to add a term of the form
\begin{equation}
  \label{eq:spurion}
  \Ltext{spurion} = \intd{\diff{^4x}} \intd{\diff{^4 \theta}} |M(m)|^2 \Psi_0
  \bar{\Psi}_0\ .
\end{equation}
Equation \eqref{eq:spurion} consistently modifies the mass of $\varphi$ as well as the minimum
condition \eqref{eq:phipotconstr}. On the other hand the second variation with respect to $\psi$
remains unchanged and thus the (pseudo-)Goldstino receives a mass. To our
present knowledge the fact that the effective spurion field must be
non-holomorphic cannot be obtained in the Wilsonian action.

Of course the realization of all symmetries should be studied carefully for
this extension. This essentially includes effects from local coupling
constants and non-perturbative effects beyond the instanton-approximation
(i.e.\ beyond the NSVZ $\beta$-function). It is beyond the scope of this paper
to answer all these questions, but it should be noted that interesting new
constraints on the general form of the effective action can be obtained by
this means: Recent results from perturbation theory showed that local couplings
in SYM theories lead to a new anomaly
\cite{kraus01,kraus01:3,Kraus:2002nu}. This anomaly is closely related to the
breaking of holomorphic dependence by higher order quantum effects. On the
other hand we expect that exactly these effects on the non-perturbative level
are responsible for dynamical supersymmetry breaking: Most calculations in
non-perturbative supersymmetry are performed in the infrared regularized
region, where holomorphic dependence is restored by the Wilsonian 1-loop
$\beta$-function. The non-holomorphic terms are obtained by a formal switch
from the Wilson action to the (infrared regularized) quantum effective
action. But neither the Wilson nor the non-holomorphic (NSVZ) $\beta$-function is
infrared stable (notice that they are both pertubative $\beta$-
functions). Thus an acceptable infrared behavior essentially includes
corrections to the $\beta$-function that could be related to an extension of the above
mentioned symmetry properties to the non-perturbative region.

%%% Local Variables: 
%%% mode: latex
%%% TeX-master: "main"
%%% End: 

%% file: conclusions.tex
\section{Conclusions}
\label{sec:conclusion}
In this paper we have introduced the extension of the Veneziano-Yankielowicz
effective action to a non-local action. Our modification affects the
non-holomorphic parts (K\"{a}hler potential) of the action only: While the
K\"{a}hler metric from the Veneziano-Yankielowicz
effective action is solely a function of the gluino-condensate, our
K\"{a}hler potential includes terms in the gluon-condensate (the ``auxiliary''
field $F$ of the effective superfield) as well. As a consequence the
effective potential depends on infinitely many coupling constants related to arbitrary powers in $\varphi$ and $F$. As the
superpotential is unchanged compared to the Veneziano-Yankielowicz action, our
extension is complementary to recent investigations on the holomorphic terms
that build on the work by Dijkgraaf and Vafa
\cite{Dijkgraaf:2002fc,Dijkgraaf:2002vw,Dijkgraaf:2002dh,Dijkgraaf:2002xd}.

The main characteristics of our effective action are:
\begin{enumerate}
\item The ``auxiliary'' field $F$ is promoted to an independent physical
  field. To obtain the spectrum of the theory it is not getting
  eliminated. Consequently, the spectrum of our model contains two complex
  scalars $\varphi$ and $F$ and one spinor $\psi$. $F$ is interpreted with the
  glue-ball and the new dynamical effects represent the consequence of the
  formation of a massive spectrum in the boson sector.
\item To arrive at a consistent formulation of point 1, the ``upside-down''
  potential for $F$ in the Veneziano-Yankielowicz Lagrangian $V_F \propto - F
  \bar{F}$ has to be replaced. The correct potential must have a minimum in $F$
  instead of a maximum. Technically this is done by formulating the effective
  potential in terms of the two fields $\Psi_0 = (\Phi)^{\inv{3}}$ and $\Psi_1
  = \bar{D}^2 \bar{\Psi}_0$ ($\Phi$ denotes the effective glue-ball superfield).
\item The spectrum of the theory does not follow from the superpotential, but
  from the generalized K\"{a}hler potential. The superpotential as a
  holomorphic function in the fields never defines stable couplings in the
  complex fields. If the highest components of the chiral fields are
  auxiliary, stable couplings are obtained after eilimination of these
  fields. As our action does not allow this elimination, the K\"{a}hler
  potential must adopt the role of the superpotential. This is possible due to
  its the rich structure. The superpotential
  reduces to the generator of the anomalies (WZ-term) and has no dominant
  influence on the spectrum.
\item Supersymmetry breaks dynamically. There are three different ways to
  establish this result:
  \begin{enumerate}
  \item A strict proof of the statement follows from the symmetries of the
    effective potential: The first variation of the effective potential with respect to
    $\varphi$ is equivalent to the mass of $\psi$. Thus the mass of $\psi$ is
    identically zero in the minimum and must be interpreted as a Goldstino
    mode.
  \item The Veneziano-Yankielowicz Lagrangian has a supersymmetric maximum in
    the ``auxiliary''-field potential (the supersymmetric vacua in the
    interpretation of ref.\ \cite{veneziano82}). As the  Veneziano-Yankielowicz Lagrangian is complete up
    to order $\bar{F} F$ in the ``auxiliary''-field potential, the maximum cannot
    be turned into a minimum. Thus the minimum must be at a different
    value of $F$, $F_0 \neq 0$.
  \item $F_0 \neq 0$ plays an important role in the dynamics. Due to this
    vacuum expectation value, new terms including explicit space-time
    derivatives follow. These terms can be important to stabilize
    the dynamics around the correct ground-state.
  \end{enumerate}
  The breaking-mechanism is of essentially non-perturbative nature and is not
  excluded by different contrarian results (Instanton calculations, Witten
  index etc.).
\item Any ground-state of the effective potential can be endowed with
  stable dynamics. This includes terms of arbitrary order in the momentum
  (non-local effective action).
\end{enumerate}

Though we are convinced that our effective action can answer several open
questions about the non-perturbative dynamics of SYM theories different
problems remain. First we have addressed in this work the invariant
terms of the action, only. Especially when considering non-vanishing values of
the source-extension this is not sufficient. Thus we should include the
complete (anomalous) symmetry structure of the effective action for local
couplings as well as for soft supersymmetry breaking. Recent results from
perturbation theory \cite{kraus01,kraus01:3,Kraus:2002nu} suggest that this
extension is closely related to the breaking of holomorphic dependence by
quantum corrections. As the impact of non-holomorphic contributions to the
potential is certainly a main result of our calculations,  a
better understanding of these terms could be of main importance. Finally, we
have shown that the semi-classical analysis of SYM theories as instanton
calculations deal with an expansion around an unstable state. Classical field
configurations responsible for the correct ground-state are unknown, but these field
configurations should be related to the glue-ball dynamics. It would be very
interesting to investigate this questions, which could be helpful to
understand the non-perturbative dynamics of non-supersymemtric gauge-theories
as well.

%%% Variables: 
%%% mode: latex
%%% TeX-master: "main"
%%% End: 